\documentclass[12pt,reqno]{amsart}

\usepackage{amsfonts}
\usepackage{amsthm}
\usepackage{mathrsfs}
\usepackage{graphicx}
\usepackage{subfigure}
\usepackage{bbm}
\usepackage{color}
\usepackage{fancyhdr}
\usepackage[latin9]{inputenc}
\usepackage{units}
\usepackage{amsbsy}
\usepackage{amstext}
\usepackage{amssymb}
\usepackage[unicode=true,
 bookmarks=false,
 breaklinks=false,pdfborder={0 0 1},colorlinks=true]
 {hyperref}
\hypersetup{
 linkcolor=myblue,citecolor=myviolet}
\usepackage{breakurl}

\makeatletter

\numberwithin{equation}{section}
\numberwithin{figure}{section}

\usepackage{color}
\definecolor{myblue}{rgb}{0.3, 0.0, 0.85}
\definecolor{myviolet}{rgb}{0.5, 0.0, 0.5}

\theoremstyle{plain}
\newtheorem{defn}{Definition}[section]
\newtheorem{Proposition}[defn]{Proposition}

\newtheorem{Theorem}[defn]{Theorem}
\newtheorem{Corollary}[defn]{Corollary}
\newtheorem{Remark}[defn]{Remark}

\title[Nodal structure of nonlinear waves on star graphs]{On the nodal structure of nonlinear stationary waves on star graphs}
\author{Ram Band}
\address{Department of Mathematics, Technion -- Israel Institute of Technology, Haifa 3200003, Israel}
\email{ramband@technion.ac.il}

\author{Sven Gnutzmann}
\address{School of Mathematical Sciences, University of Nottingham,
  Nottingham NG7 2RD,  United Kingdom}
\email{sven.gnutzmann@nottingham.ac.uk}

\author{August J. Krueger}
\address{Department of Mathematics,
  Rutgers University,
  Piscataway, NJ 08854-8019, USA}
\email{akrueger@math.rutgers.edu}

\date{\today}

\makeatother

\begin{document}
\maketitle
\begin{abstract}
  We consider stationary waves on nonlinear quantum star
    graphs, i.e.
    solutions to the stationary (cubic) nonlinear Schrödinger equation on a
    metric star graph with Kirchhoff matching conditions at the centre.
    We prove the existence of solutions that vanish at the centre of the star
    and classify them according to the nodal structure on each edge
    (i.e. the number of nodal domains or nodal points that the solution
    has on each edge). We discuss the relevance of these solutions in more
    applied settings as starting points for numerical calculations of spectral
    curves and put our results into the wider context of nodal counting such as
    the classic Sturm oscillation theorem.
\end{abstract}

\maketitle

\section{Introduction}

\label{sec:introduction}

Sturm's oscillation theorem \cite{Sturm1836} is a classic example for how
solutions of linear self-adjoint differential eigenvalue problems
$D\phi(x)=\lambda\phi(x)$ (where $D$ is a Sturm-Liouville operator) are ordered and classified by the number of nodal
points. According to Sturm's oscillation theorem, the $n$-th eigenfunction, $\phi_{n}$, has
$n-1$ nodal points, when the eigenfunctions are ordered by increasing order of their corresponding eigenvalues
$\lambda_{1}<\lambda_{2}<\lambda_{3}<\dots$. Equivalently,
the number $\nu_{n}$ of nodal domains (the connected domains where
$\phi_{n}$ has the same sign) obeys $\nu_{n}=n$ for all $n$.

In higher dimensions, (e.g., for the free Schr{\"o}dinger equation $-\Delta\phi(\mathbf{x})=\lambda\phi(\mathbf{x})$
on a bounded domain with self-adjoint boundary conditions) the number
of nodal domains is bounded from above, $\nu_n\le n$, by Courant's
theorem \cite{Courant23} (see \cite{AncHelfHof_doc04} for the case of Schr{\"o}dinger equation with potential). Furthermore, there is only a finite number of Courant
sharp eigenfunctions for which $\nu_{n}=n$, as was shown by Pleijel
\cite{Pleijel56}.

In (linear) quantum graph theory one considers the Schr{\"o}dinger equation
with self-adjoint matching conditions at the vertices of a metric
graph. Locally, graphs are one-dimensional though the connectivity of
the graph allows to mimic some features of higher dimensions.
Nodal counts for quantum graphs have been considered
for more than a decade \cite{GnuSmiWeb_wrm04}.
For example, it has been shown in \cite{GnuSmiWeb_wrm04} that Courant's
bound applies to quantum graphs as well. Yet, for graphs there are
generically infinitely many Courant sharp eigenfunctions \cite{BanBerWey_jmp15, AloBanBer_cmp18}.
For tree graphs it has been proven that all generic eigenfunctions are Courant sharp, i.e.,
$\nu_{n}=n$ \cite{PokPryObe_mz96,Schapotschnikow06} .
In other words, Sturm's oscillation theorem generalizes to metric trees graphs. It has been proven
by one of us that the converse also holds, namely that if the graph's
nodal count obeys $\nu_{n}=n$ for all $n$, then this graph is a tree
\cite{Ban_ptrsa14}.
When a graph is not a tree, its first Betti number, $\beta:=E-V+1$ is positive.
Here, $E, V$ are correspondingly the numbers of graph's edges and vertices
and $\beta$ indicates the number of the graphs independent cycles.
In addition to Courant's bound, the nodal count of a graph is bounded from
below, $\nu_n\leq n-\beta$ as was shown first in \cite{Berkolaiko07}.
The actual number of nodal domains may be characterized by various variational
methods~\cite{BanBerRazSmi_cmp12, BerWey_ptrsa13}.
Some statistical properties of the nodal count are also
known \cite{AloBanBer_cmp18}, but to date there is no general
explicit formula or a full statistical description of the nodal count.

\begin{figure}[h]
  \centering
  \includegraphics[width=0.4\textwidth]{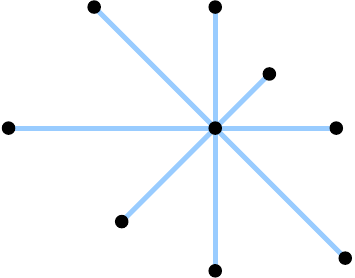}
  \caption{\label{fig:star} A star graph with $E=8$ edges and $E+1=9$ vertices.}
\end{figure}

In the present work we present some related results concerning nodal
points and nodal domains for \emph{nonlinear} star graphs (see Figure~\ref{fig:star}).  
Nonlinear wave equations on metric graphs (i.e., nonlinear
quantum graphs) have recently attracted considerable interest both
from the mathematical perspective and the applied regime. They allow
the study of intricate interplay between the non-trivial connectivity and the
nonlinearity. Among the physical applications of nonlinear wave equations
on metric graphs is light transmission through a network of optical fibres or
Bose-Einstein condensates in quasi one-dimensional traps.
We refer to \cite{GW,GW2} where a detailed overview of the recent
literature and some applications
is given and just summarise here the relevant work related to
the nodal counting.
In a previous
work some of us have shown that Sturm's oscillation theorem is generically
broken for nonlinear quantum stars, apart from the special case of
an interval \cite{BanKru_amspro18}. This is not unexpected as the
set of solutions
is known to have a far more complex structure. Our main result here
is that the nonlinear case of a metric star allows for solutions with any given
number of nodal domains on each edge. Namely, for a star with $E$ edges and
a certain $E$-tuple, $(n_{1},\dots,n_{E})$ of
non-negative integers there are solutions with $n_{e}$ nodal points
on the $e$-th edge
for $e=1,\, \dots,E$.

In the remainder of the introduction chapter we define the setting.
In Section~\ref{sec:theorems}
we state our main results. In Section~\ref{sec:background}
we present the nonlinear generalization of Sturm's oscillation theorem
to an~interval, some general background and properties of
nonlinear solutions as well as a few motivating numerical
results. In Section \ref{sec:proofs} we prove the main theorems and
afterwards in Section~\ref{sec:discussion}
we discuss our results and their possible implications in the
broader context of nonlinear quantum graphs.

\subsection{The Setting---Nonlinear Star Graphs}

\label{sec:setting}

Metric star graphs are a special class of metric trees with $E$ edges and $E+1$ vertices such that all edges are incident
to one common vertex (see Figure~\ref{fig:star}). The common vertex will be called the centre
of the star and the other vertices will be called the boundary. We
assume that each edge has a~finite length $0<\ell_{e}<\infty$ ($e=1,\dots,E$)
and a coordinate $x_{e}\in[0,\ell_{e}]$ such that $x_{e}=0$ at the
centre and $x_{e}=\ell_{e}$ at the boundary. On each edge $e=1,\dots,E$
we consider the stationary nonlinear Schr{\"o}dinger (NLS)~equation
\vspace{6pt}
\begin{equation}
-\frac{d^{2}}{dx_{e}^{2}}\phi_{e}(x_{e})+g|\phi_{e}(x_{e})|^{2}\phi_{e}(x_{e})=\mu\ \phi_{e}(x_{e})
\label{eq:NLSE}
\end{equation}
for $\phi_{e}\, : \, [0,\ell_{e}] \rightarrow \mathbb{C} $.
Here $g$ is a nonlinear coupling parameter and $\mu$ a spectral parameter.
We consider this as a generalized eigenequation with eigenvalues $\mu$.
We have assumed here that the nonlinear interaction is homogeneously
repulsive ($g>0$) or attractive ($g<0$) on all edges and will continue to do so throughout
this manuscript.
One may consider more general graphs
where $g$ takes different values (and different signs) on different
edges (or even where $g\rightarrow g_e(x_e)$ is a real scalar function on
the graph).
It is not, however, our aim to be as general as possible. In
the following we restrict ourselves to one generic setting in order
to keep the notation and discussion as clear and short as possible.
We~will later discuss some straightforward generalizations of our
results.

At the centre we prescribe Kirchhoff (a.k.a Neumann) matching conditions
\begin{align}
\phi_{e}(0)= & \phi_{e'}(0)\quad & \text{for all \ensuremath{1\le e<e'\le E}}\label{eq:Kirchhoff-1}\\
\sum_{e=1}^{E}\frac{d\phi_{e}}{dx_{e}}(0)= & 0\label{eq:Kirchhoff-2}
\end{align}
and at the boundary we prescribe Dirichlet conditions $\phi_{e}(\ell_{e})=0$.

For the coupling constant $g$ it is sufficient without loss of
generality to consider three different~cases. For~$g=0$ one recovers the linear Schr{\"o}dinger equation
and thus standard quantum star graphs. For~$g=1$ one has a nonlinear
quantum star graph with repulsive interaction and for $g=-1$ one
has a nonlinear quantum star with attractive interaction.
If $g$ takes any other non-zero value a simple rescaling
  $\phi_e(x_e) \mapsto \frac{1}{\sqrt{|g|}}\phi_e(x_e)$ of the
    wavefunction is equivalent to replacing $g \mapsto \frac{g}{|g|}
    =\pm 1$.

Without loss of generality we may focus on real-valued
and twice differentiable solutions $\{\phi_{e}(x_{e})\}_{e=1}^{E}$,
where twice differentiable refers separately to each $\phi_{e} : (0,\ell_{e}) \rightarrow \mathbb{C} $.
We also assume that the solution is not the constant zero function on the graph, namely that there is an edge $e$ and some point $\hat{x}_{e}\in[0,\ell_{e}]$
with $\phi_{e}(\hat{x}_{e})\neq0$. Moreover, any complex-valued solution
is related to a real-valued solution by a global gauge-transformation
(i.e., a change of phase $\phi_{e}(x_{e})\mapsto\phi_{e}(x_{e})e^{i\alpha}$)
\cite{GW}.

\subsection{The Nodal Structure}

\label{sec:intro_nodal}

We will call a solution $\{\phi_{e}(x_{e})\}_{e=1}^{N}$ \emph{regular}
if
the wavefunction does not vanish on any edge,
that~is for each
edge $e$ there is $\hat{x}_{e}\in(0,\ell_{e})$ with
$\phi_{e}(\hat{x}_{e})\neq 0$.
Accordingly, \emph{non-regular} solutions vanish identically on some
edges, in other words there is (at least) one edge $e$ such that
$\phi_{e}(x_{e})=0$ for all $x_{e}\in[0,\ell_{e}]$.

A solution with a node at the centre, $\phi_{e}(0)=0$ (by continuity
this is either true for all $e$ or for none) will be called \emph{central
Dirichlet} because it satisfies Dirichlet conditions at the centre
(\mbox{in addition to} the Kirchhoff condition). Hence, non-regular solutions are always
central Dirichlet. Our main theorem will construct solutions which
are regular and central Dirichlet. Note that from a regular central
Dirichlet solution on a metric star graph $G$ one can construct non-regular
solutions on a larger metric star graph~$G'$, if $G$ is a metric
subgraph of $G'$: on each edge $e\in G'\setminus G$ one may just
extend the solution by setting $\phi_{e}(x_{e})=0$ for all $x_{e}\in[0,\ell_{e}]$.

Our main aim is to characterize solutions in terms of their nodal
structure. The nodal structure is described in terms of either the number $\nu$ of nodal domains (maximal connected
subgraphs where $\phi_{e}(x_{e})\neq0$) or by the number $\xi$
of nodal points. We will include in the count the trivial nodal points at the boundary.
Note that regular solutions which are not central
Dirichlet obey $\nu=\xi+1-E$ while regular central Dirichlet solutions
obey $\nu=\xi-1$. We have stated in the introduction that in the
linear~case, $g=0$, such a characterization is very well understood
even for the more general tree~graphs, which obey a generalized version of Sturm's oscillation theorem.

As we will see, the solutions of nonlinear star graphs have a very
rich structure and a classification of solutions in terms of the total
numbers $\nu$ or $\xi$ of nodal domains or nodal points is far from
being unique. We will thus use a more detailed description of the nodal
structure of the solutions.
 To each regular solution $\{\phi_{e}(x_{e})\}_{e=1}^{E}$ we associate
the $E$-tuple
\begin{equation}
\mathbf{n}=(n_{1},\dots,n_{E})\in\mathbb{N}^{E}
\end{equation}
where $n_{e}\ge1$ is the number of nodal domains of the wavefunction
$\phi_{e}(x_{e})$ on the edge $x_{e}\in[0,\ell_{E}]$. For~solutions
which are not central Dirichlet, $n_{e}$ also equals the number of nodal points of $\phi_{e}(x_{e})$ (including the nodal point at the boundary).
We will call $\mathbf{n}\in\mathbb{N}^{E}$ the \emph{(regular) nodal
edge count structure} of the (regular) solution $\{\phi_{e}(x_{e})\}_{e=1}^{N}$.
For non-regular solutions one may characterize the nodal structure
in a similar way by formally setting $n_{e}=\infty$ for all edges
where the wavefunction is identical~zero. In that case we speak of
a \emph{non-regular nodal edge count structure}.
Note that we do not claim that the nodal edge count structure, $\mathbf{n}$, leads to a unique
  characterization of the solutions (which actually come in
  one-parameter families). Indeed we have numerical counter-examples.
  With this more detailed description we show that a much larger set of nodal
  structures is possible in nonlinear quantum star graphs compared to
  the linear case, as is stated in the next section.

\section{Statement of Main Theorems}

\label{sec:theorems}

Our main results concern the existence of solutions with any
given nodal edge structure. We
state two theorems: One
for repulsive nonlinear interaction $g=1$ and one for attractive
nonlinear interaction $g=-1$.
The two
theorems establish the existence of central Dirichlet solutions with
nodal edge structure $\mathbf{n}=(1,\ldots,1)$ subject to (achievable)
conditions on the edge lengths. As corollaries, we
get the existence of central Dirichlet solutions with any prescribed
values of $\mathbf{n}$ (again subject to some
  achievable conditions on the lengths).
Throughout this section we consider a nonlinear
quantum star graph as described in Section~\ref{sec:setting}.
In order to avoid trivial special cases we will assume $E\ge 3$. Indeed,
$E=1$ is the interval and well understood and $E=2$ reduces to an
interval (of total length $\ell_{1}+\ell_{2}$) as the Kirchhoff vertex
condition in this case just states that the wavefunction is continuous
and has a continuous first derivative. We will also assume that all
edge lengths are different. Without loss of generality we take them
as ordered $\ell_{e}<\ell_{e+1}$ ($e=1,\dots,E-1$).

\begin{figure}[h]
\centering
  \includegraphics[width=0.75\textwidth]{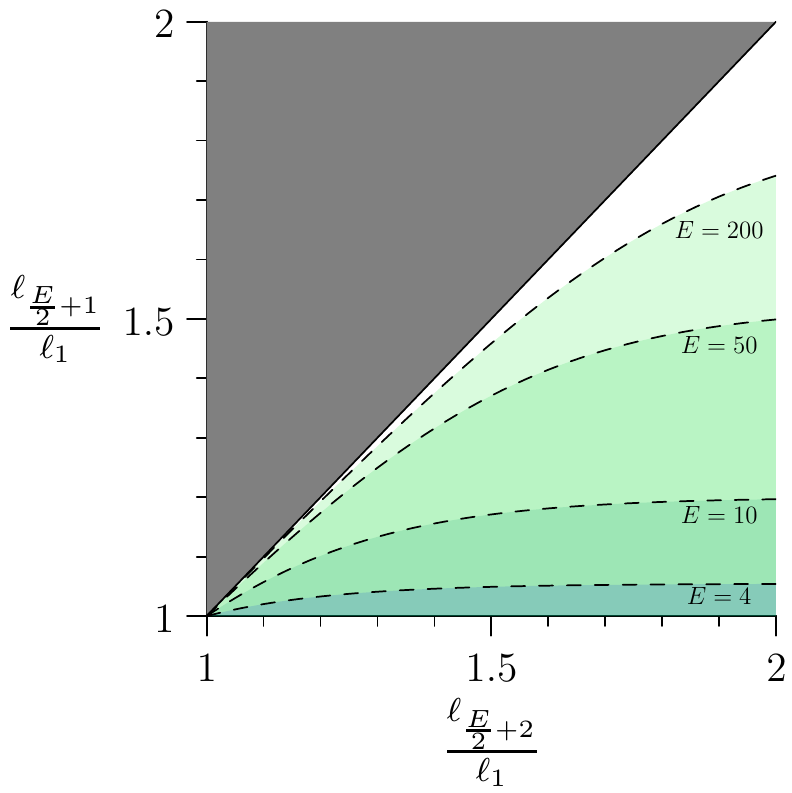}
  \caption{\label{fig1}
    The shaded regions indicate
    choices of relative
    edge lengths $1< \frac{\ell_{\frac{E}{2}+1}}{\ell_1}<
    \frac{\ell_{\frac{E}{2}+2}}{\ell_1}$
    that satisfy the condition \eqref{cond_thm1} of
    Theorem~\ref{thm1}.
    The dashed lines indicate the boundary of
    regions for a star graph
    with $E$ edges (where $E=4, 10, 50, 200$).
    Condition \eqref{cond_thm1} is
    satisfied below the dashed lines.
  }
\end{figure}

\begin{Theorem} \label{thm1} If $g=1$ (repulsive case) and either
  \begin{enumerate}
  \item the number of edges $E$ is odd, or
  \item $E$ is even and
    \begin{equation}
      \sqrt{\frac{m_{+}}{m_{-}}}\frac{1+m_{-}}{1+m_{+}}>\frac{E}{E-2},
      \label{cond_thm1}
    \end{equation}
     where $0 < m_{-} < m_{+} < 1$ are implicitly defined
    in
      terms of the edge lengths $l_{1},~l_{\frac{E}{2}+1},~l_{\frac{E}{2}+2}$
    by
    \begin{equation}
      \begin{split}
        K(m_{+})\sqrt{1+m_{+}}= & \frac{\pi}{2}\,\frac{\ell_{\frac{E}{2}+2}}{\ell_{1}},\\
        K(m_{-})\sqrt{1+m_{-}}= & \frac{\pi}{2}\,\frac{\ell_{\frac{E}{2}+1}}{\ell_{1}},
      \end{split}
      \label{eq:thm1_mplusminus}
    \end{equation}
    with
    \begin{equation}
    K(m)=\int_{0}^{1}\frac{1}{\sqrt{1-u^{2}}\sqrt{1-mu^{2}}}du
    \label{eq:K_m}
    \end{equation}
    being the complete elliptic integral of first kind,
  \end{enumerate}
  then there exists a regular central Dirichlet solution for some
  positive value of the spectral parameter $\mu=k^{2}> \frac{\pi^2}{\ell_1^2}$
  such that there is exactly
  one nodal domain on each edge,
  i.e., the nodal
  edge structure $\mathbf{n}$
  satisfies $n_{e}=1$ for all edges $e$.
\end{Theorem}
Note that the condition in this theorem for even number of edges
involves only three edge
lengths
and can be stated in terms of two ratios that satisfy
$\frac{\ell_{\frac{E}{2}+2}}{\ell_{1}}\ge
\frac{\ell_{\frac{E}{2}+1}}{\ell_{1}}\ge 1$ (as we have ordered the
edges by lengths). If the larger ratio
$\frac{\ell_{\frac{E}{2}+2}}{\ell_{1}}$
is given then one may always achieve this condition by choosing the
other
ratio sufficiently small (as
$\frac{\ell_{\frac{E}{2}+1}}{\ell_{1}}\to 1$ one has $m_-\to 0$ and
the left-hand side of condition \eqref{cond_thm1} grows without any
bound). Figure~\ref{fig1} shows a graph of the regions
where the two length ratios satisfy condition \eqref{cond_thm1} for
star graphs with $E$ edges. One can see how the condition becomes
less restrictive when the number of edges is large. We will present
the proof of Theorem~\ref{thm1} in Section~\ref{sec:proof_thm1}.
The proof shows that the condition \eqref{cond_thm1} is not optimal.
Less restrictive conditions that depend on other edge lengths may be stated.
Nevertheless, we have chosen to state the condition \eqref{cond_thm1}, as its form
is probably more compactly phrased than other conditions would be.

\begin{Theorem}
  \label{thm2}
  Let $g=-1$ (attractive case).
  If there exists an integer
  $M<E/2$
  such that
  \begin{equation}
    \sum_{e=M+1}^{E-1}\frac{1}{\ell_{e}^{2}}<\sum_{e=1}^{M}\frac{1}{\ell_{e}^{2}}<\sum_{e=M+1}^{E}\frac{1}{\ell_{e}^{2}}, \label{cond1_thm2}
  \end{equation}
  then there exists a regular central Dirichlet solution for some
  positive value of the spectral parameter $\mu=k^{2}\in
  \left(0,\frac{\pi^2}{\ell_{E}^2}\right)$ such that there is
  exactly
  one nodal domain on each edge, i.e., the
  nodal edge structure $\mathbf{n}$ satisfies $n_{e}=1$ for all edges
  $e$.
\end{Theorem}

We will prove this theorem in Section~\ref{sec:proof_thm2}.
One may extend Theorem~\ref{thm2} to find negative values of the spectral
parameter $\mu<0$ under appropriate conditions on edge lengths using
similar ideas as the ones used in our proof for $\mu>0$. To keep the
paper concise we focus here on $\mu>0$.

In order to demonstrate how the two conditions \eqref{cond1_thm2} in Theorem \ref{thm2} may be achieved, we point out that the following weaker conditions
\begin{equation}
  \begin{split}
     \frac{\ell_M}{\ell_{M+1}}<&\sqrt{\frac{M}{E-M-1}} \\
    \frac{\ell_1}{\ell_E}>&\sqrt{\frac{M}{E-M}}
  \end{split}
  \label{cond2_thm2}
\end{equation}
imply \eqref{cond1_thm2} (recalling that $\ell_1<\ldots<\ell_E$). The conditions \eqref{cond2_thm2}
are easy to apply and they may be achieved
straight-forwardly.
For instance, if $E$ is odd and $M=(E-1)/2$
(the largest possible value for $M$) then  the first inequality in
\eqref{cond2_thm2} is always satisfied and the second condition
gives the restriction $1>  \frac{\ell_1}{\ell_E}>
\sqrt{\frac{M}{M+1}}$ on the ratio between the smallest and largest
edge length. In addition to that one may easily construct a star graph
with edge lengths which satisfy conditions \eqref{cond2_thm2} above.
This is done by starting from a star graph which has
only two different edge lengths $\ell_-< \ell_+$ where
$\ell_e=\ell_-$
for $1\le e\le M$ and $\ell_e=\ell_+$ for $M+1\le e \le E$.
If one chooses the ratio of the lengths in the range
$\sqrt{\frac{M}{E-M}} < \frac{\ell_-}{\ell_+}<
\sqrt{\frac{M}{E-M-1}}$
and then perturbs all edge lengths slightly to make them different
then
condition \eqref{cond2_thm2} is satisfied. Note however that, just
as in Theorem~\ref{thm1}, even the condition which is stated in Theorem~\ref{thm2}
is not optimal and more detailed conditions can be derived from our
proof in Section~\ref{sec:proof_thm2}.

Before discussing some straight-forward implications
let us also state here that the assumption that all edge lengths are
different that we made for both Theorems~\ref{thm1} and~\ref{thm2} may be relaxed.
This is because any two edges with the same length decouple
in a certain way from the remaining graph. If~one deletes pairs of
edges of equal length from the graph until all edges in the remaining
graph are different one may apply the theorems to the remaining graph
(if the remaining graph has at least three edges).
This will be discussed more
in Remark~\ref{rem: Equal-edge-lengths}.

In the remainder of this section we discuss
the implications of the two theorems for finding solutions with
a given nodal edge structure $\mathbf{n}\in \mathbb{N}^E$.
In this case we divide each edge length into $n_e$ fractions $\ell_e= n_e
\tilde{\ell}_e$. The $n$-th fraction $\tilde{\ell}_e$ then corresponds to the
length of one nodal domain. For the rest of this section we do not
assume
that the edge lengths $\{\ell_e\}$ are ordered by length and different,
rather~we now
assume that these assumptions apply to the fractions, i.e.,
$\tilde{\ell}_e<\tilde{\ell_{e+1}}$ ($e=1,\dots,E-1$).
By~first considering the metrically smaller star graph with edge
lengths $\{\tilde{\ell}_e\}$ Theorems~\ref{thm1} and \ref{thm2}
establish the existence of solutions on this smaller graph subject
to conditions on the lengths $\{\tilde{\ell}_e\}$. These solutions
can be extended straight-forwardly to a solution on the full star
graph. Indeed, as we explain in more detail in Section~\ref{sec:interval}, the
solution on each edge is a naturally periodic function given by an elliptic deformation
of a sine and shares the same symmetry around nodes and extrema, as the sine function.
The main relevant difference to a sine is that the period of the
solution depends on the amplitude. In the repulsive case one then obtains
the following.


\begin{Corollary} \label{cor1}
    Let $g=1$ (repulsive case) and $\mathbf{n}\in N^{E}$.
    If either
    \begin{enumerate}
    \item $E$ is odd, or
    \item $E$ is even and the fractions $\tilde{\ell}_e=\ell_e/n_e$
      ($e=1,\dots, E$) satisfy the condition \eqref{cond_thm1},
    \end{enumerate}
    then there exists a regular central Dirichlet solution for
    some positive value of the spectral
    parameter $\mu=k^{2}>\frac{\pi^2}{\tilde{\ell}_1^2}$
    `with regular nodal edge count structure $\mathbf{n}$.
\end{Corollary}

Similarly,
Theorem~\ref{thm2} implies the following.

\begin{Corollary} \label{cor2}
    Let $g=-1$ (attractive case) and  $\mathbf{n}\in\mathbb{N}^{E}$.
    If the fractions $\tilde{\ell}_e=\ell_e/n_e$ satisfy condition
    \eqref{cond1_thm2}
    then there exists a regular central Dirichlet solution for
    some positive value of the spectral
    parameter $\mu=k^{2}\in \left(0, \frac{\pi^2}{
         {\tilde{\ell}_E}^{2} } \right)$
    with regular nodal edge count structure $\mathbf{n}$.
\end{Corollary}

The corollaries above provide
sufficient conditions for the existence of a central Dirichlet solution
with a particular given nodal edge count. In addition to that, it
is straight-forward to apply
Theorems~\ref{thm1} and \ref{thm2}
to show that for any choice
of edge lengths there are infinitely many $E$-tuples which can serve
as the graph's regular central Dirichlet nodal structure.

Moreover Theorems~\ref{thm1} and \ref{thm2} also imply infinitely
many values for
non-regular nodal
structures,
as every non-regular solution is equivalent to a regular solution
on a subgraph.

Finally, we note that the proofs of
Theorems~\ref{thm1} and \ref{thm2}
in Section~\ref{sec:proofs}
are constructive and they specify the corresponding solution
up to a single parameter (which one may
take to be $k=\sqrt{\mu}$) that may easily be found numerically.

\section{General Background on the Solutions of Nonlinear
  Quantum Star
  Graphs}

\label{sec:background}

Before we turn to the proof of Theorems~\ref{thm1} and \ref{thm2}
we would like discuss how the implied regular central Dirichlet solutions
are related to the complete set of solutions of the nonlinear star
graph. Though we are far from having a full understanding of all solutions
we can give a heuristic picture.

\subsection{The Nonlinear Interval - Solutions and Spectral Curves}

\label{sec:interval}

Let us start with giving a complete overview of the solutions for
the interval (i.e., the star graph with $E=1$). While these are well
known and understood they play a central part in the construction
of central Dirichlet solutions for star graphs in our later proof
and serve as a good way to introduce some general background. On the
half line $x\ge0$ with a Dirichlet condition $\phi(0)=0$ at
the origin it is straight-forward to check (see also \cite{GW}) that the solutions for positive spectral parameters
$\mu=k^{2}$ (where $k>0$) are of the form
\begin{equation}
  \phi(x)=
  \begin{cases}
    \chi_{m,k}^{(+)}(x)=
    k\sqrt{\frac{2m}{1+m}}\ \mathrm{sn}\left(\frac{kx}{\sqrt{1+m}},m\right) &
    \text{in the repulsive case \ensuremath{g=1} ,}\\[0.2cm]
    \chi_{m,k}^{(-)}(x)=
    k\sqrt{\frac{2m(1-m)}{1-2m}}\ \frac{\mathrm{sn}\left(\frac{kx}{\sqrt{1-2m}},m\right)}{\mathrm{dn}\left(\frac{kx}{\sqrt{1-2m}},m\right)} &
    \text{in the attractive case \ensuremath{g=-1}.}
  \end{cases}
  \label{chi_sol}
\end{equation}
Here $\mathrm{sn}(y,m)$ and $\mathrm{dn}(y,m)$ are Jacobi elliptic
functions with a deformation parameter $m$. The definition of elliptic
functions allows $m$ to take arbitrary values in the interval $m\in[0,1]$
(as there are many conventions for these functions we summarize ours
in Appendix~\ref{sec:elliptic}). Note that $\mathrm{sn}(y,m)$ is
a deformed variant of
the sine function and $\mathrm{sn}(y,0)=\sin(y)$ and $\mathrm{dn}(y,0)=1$.\\
 For any spectral parameter $\mu=k^{2}$ there is a one-parameter
family of solutions parameterised by the deformation parameter $m$.
In the repulsive case the deformation parameter may take values $m\in(0,1]$
(as for $m=0$ one obtains the trivial solution $\chi_{0,k}^{(+)}(x)=0$)
and in the attractive case $m\in\left(0,\frac{1}{2}\right)$ (the~expressions are not well defined for $m=\frac{1}{2}$ and for $m>\frac{1}{2}$
the expressions are no longer real).

Let us now summarise some properties of these solutions in the following
proposition for the
solutions of the NLS equation on the half line. \begin{Proposition} The
solutions $\phi(x)=\chi_{m,k}^{(\pm)}(x)$ given in Equation \eqref{chi_sol} have
the following properties
\begin{enumerate}
\item All solutions are periodic $\chi_{m,k}^{(\pm)}(x)=\chi_{m,k}^{(\pm)}(x+\Lambda^{(\pm)}(m,k))$
with a nonlinear wavelength
\begin{equation}
  \begin{split}
    \Lambda^{(+)}(m,k)=&
    \frac{4\sqrt{1+m}K(m)}{k} 
    \\[0.3cm]
    \Lambda^{(-)}(m,k)=&
    \frac{4\sqrt{1-2m}K(m)}{k}, 
  \end{split}
\end{equation}
where $K(m)$ is the complete elliptic integral of first kind, Equation~(\ref{eq:K_m}).
\item
  For $m\to 0$ one regains the standard relation $\Lambda^{(\pm)}(0,k)=\frac{2\pi}{k}$
  for the free linear Schr{\"o}dinger equation. In~the repulsive case $\Lambda^{(+)}(m,k)$
  is an increasing function of $m$ (at fixed $k$) that increases without
  bound as $m\to1$.
  In the attractive case $\Lambda^{(-)}(m,k)$ is
  a decreasing function of $m$ (at fixed $k$) with
  $\Lambda^{(-)}\left(\frac{1}{2},k\right)=0$.
\item The nodal points are separated
  by half the nonlinear wavelength. Namely,  $\chi_{m,k}^{(\pm)}(n\Lambda^{(\pm)}(m,k)/2)=0$
for $n=0,1,,\dots$.
\item The solutions are anti-symmetric around each nodal point and
  symmetric around each extremum, \mbox{i.e., it has} the same symmetry
  properties as a sine function.
\item As $\mathrm{sn}\left(K(m),m\right)=1$ and $\mathrm{dn}\left(K(m),m\right)=\sqrt{1-m}$
the amplitude
 \begin{small}
\begin{equation*}
  A^{(\pm)}(k,m)=\max\left(\chi_{m,k}^{(\pm)}(x)\right)_{x\ge0}=
  \chi_{m,k}^{(\pm)}\left(\frac{\Lambda^{(\pm)}(m,k)}{4}\right)
\end{equation*}
\end{small}
is given by
 \begin{small}
\begin{equation}
  \begin{split}
    A^{(+)}(m,k)=&
    k\sqrt{\frac{2m}{1+m}}
    \\[0.3cm]
    A^{(-)}(m,k)=&
    k\sqrt{\frac{2m}{1-2m}}. 
  \end{split}
\end{equation}
\end{small}
\item As $m\to 0^{+}$ the amplitude of the solutions also decreases to zero
  $A^{(\pm)}(0,k)=0$ for both the repulsive and the attractive case. In
  this case the effective strength of the nonlinear interaction becomes
  weaker and the oscillations are closer. In the repulsive case the amplitude remains bounded
  as $m\to1$ with $A^{(+)}(1,k)=k$. In the attractive case $A^{(-)}(m,k)$
  grows without bound as $m\to\frac{1}{2}$.
\end{enumerate}
\end{Proposition} All statements in this proposition follow straight-forwardly
from the known properties of elliptic integrals and elliptic functions
and we thus omit the proof here.
Furthermore, some of the statements in the proposition are mentioned explicitly in \cite{BanKru_amspro18, GW, NIST} and
  others follow easily from the definitions as given in the
  Appendix~\ref{sec:elliptic}.

For the NLS equation for $\phi(x)$ on an interval $x\in[0,\ell]$
with Dirichlet conditions at both boundaries $\phi(0)=\phi(\ell)=0$
one obtains a full set of solutions straight-forwardly from the solutions
$\chi^{(\pm)}_{m,k}(x)$ on the half-line by requiring that there
is a nodal point at $x=\ell$. Since the distance between two nodal
points in $\chi^{(\pm)}_{m,k}(x)$ is $\Lambda^{(\pm)}(k,m)/2$ the length of
the interval has to be an integer multiple of half the nonlinear wavelength
\begin{equation}
2\ell=n\Lambda^{(\pm)}(k,m)\label{interval_condition},
\end{equation}
where the positive integer $n$ is the number of nodal domains. We
arrive at the following proposition.
\begin{Proposition}
  The NLS Equation (\ref{eq:NLSE}) on an interval of length $\ell$ with
  Dirichlet boundary conditions \mbox{has a one-parameter} family of real-valued
  solutions with $n$ nodal domains,
  for each $n\in\mathbb{N}$. The relation between the spectral
  parameter $\mu=k^{2}$ and the deformation parameter $m$ is dictated
  by Equation~(\ref{interval_condition}) and may be explicitly written as
  \begin{equation}
    \begin{split}
      k^{(+)}_{n,\ell}(m)=
      &
      \frac{2n\sqrt{1+m}K(m)}{\ell}
      \\[0.3cm]
      k^{(-)}_{n,\ell}(m)=
      &
      \frac{2n\sqrt{1-2m}K(m)}{\ell}
    \end{split}
    \label{k_m}.
  \end{equation}
\end{Proposition}

\begin{figure}[h]
\centering
  \includegraphics[width=0.49\textwidth]{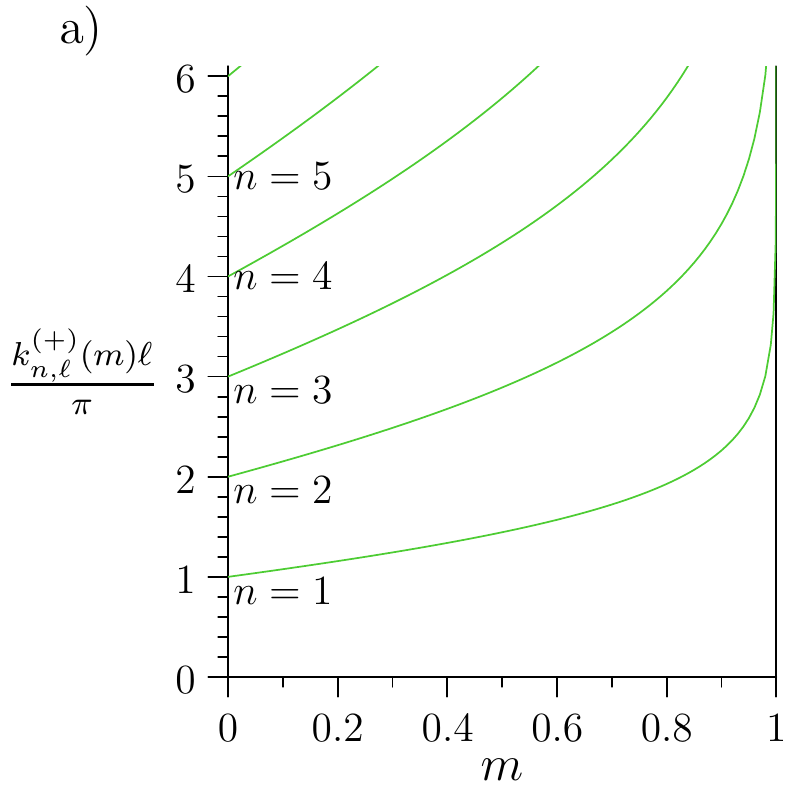}
  \includegraphics[width=0.49\textwidth]{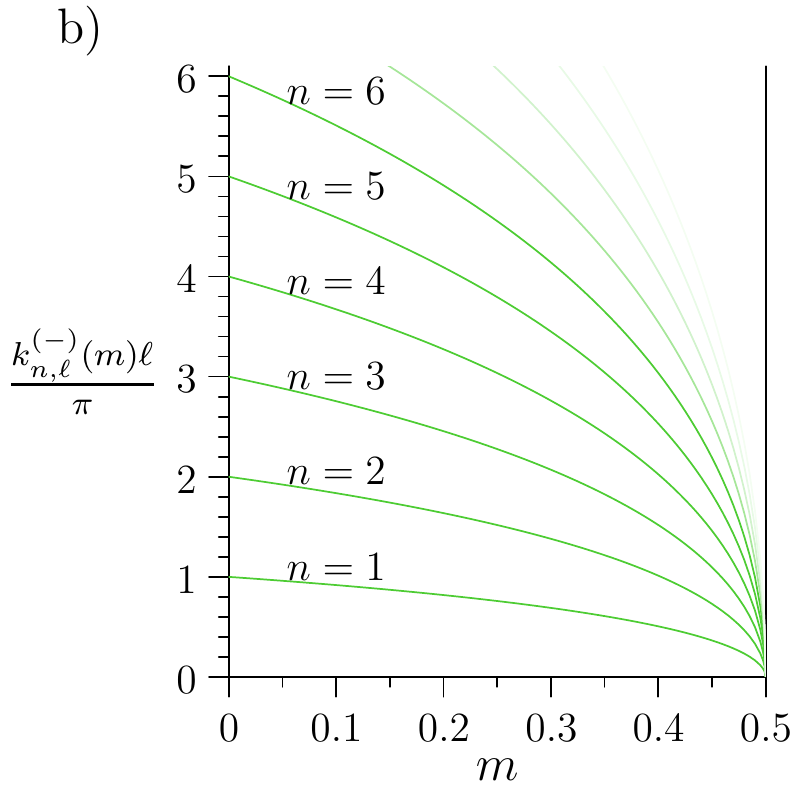}
\caption{\label{fig2} {Spectral curves
  $k^{(\pm)}_{n,\ell}(m)$  for the repulsive (\textbf{a}) and attractive case
  (\textbf{b}). The $n$-th curve is obtained from the curve for $n=1$ by
  rescaling $k^{(\pm)}_{n,\ell}(m)=n k^{(\pm)}_{1,\ell}(m)$.}
}
\end{figure}

We refer to $k^{(\pm)}_{n\ell}(m)$ (or its implicitly defined inverse
$m^{(\pm)}_{n,\ell}(k)$) as spectral curves.
As $k^{(\pm)}_{n+1,\ell}(m)>k^{(\pm)}_{n,\ell}(m)$,
the spectral curves never cross (see Figure~\ref{fig2}) and we obtain
the first nonlinear generalization of Sturm's oscillation theorem
as a corollary
(see also
Theorem 2.4 in \cite{BanKru_amspro18}).

\begin{Corollary}
  For any allowed value of the deformation parameter $m$
  ($m\in(0,1)$ for $g=1$ and $m\in\left(0,\frac{1}{2}\right)$ for
  $g=-1$) there is a discrete set $\{k_{n}\}_{n=1}^{\infty}$ of positive
  real numbers, increasingly ordered,
  such that $\phi_{n}=\left.\chi^{(\pm)}_{m,k_{n}}\right|_{[0,\ell]}$ is a solution of the NLS equation
  on the interval $[0,\ell]$ with spectral parameters $\mu_{n}=k_{n}^{2}$ and
  $n$ is the number of nodal domains.
  Furthermore, these are all the solutions of the
  NLS equation whose deformation parameter equals~$m$.
\end{Corollary}

While this is mathematically
sound,
fixing the deformation parameter $m$ is not a very useful approach
in an applied setting. A more physical approach (and one that is useful
when we consider star graphs) is to fix the $L^{2}$-norm
$N^{(\pm)}_{n,\ell}(m)=\int_{0}^{\ell}\chi^{(\pm)}_{m,k^{(\pm)}_{n,\ell}(m)}(x)^{2}dx$
of the solutions. The $L^{2}$-norm is a global measure for the strength
of the nonlinearity. It has the physical meaning of an integrated
intensity. In optical applications this is proportional to the total
physical energy and for applications in Bose-Einstein condensates
this is proportional to the number of particles.

\begin{figure}[h]
\centering
  \includegraphics[width=0.46\textwidth]{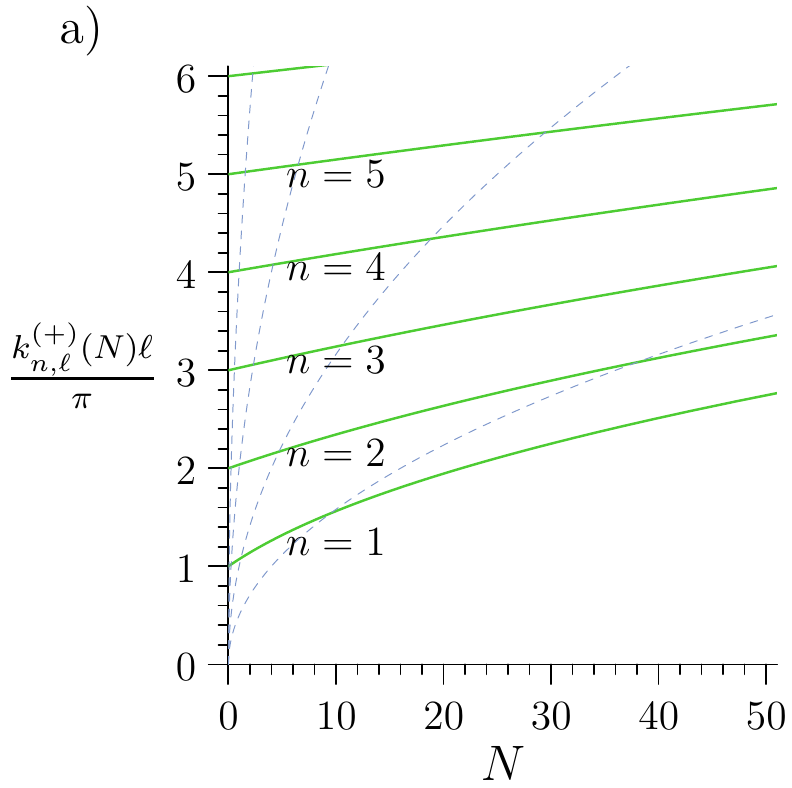}
  \includegraphics[width=0.46\textwidth]{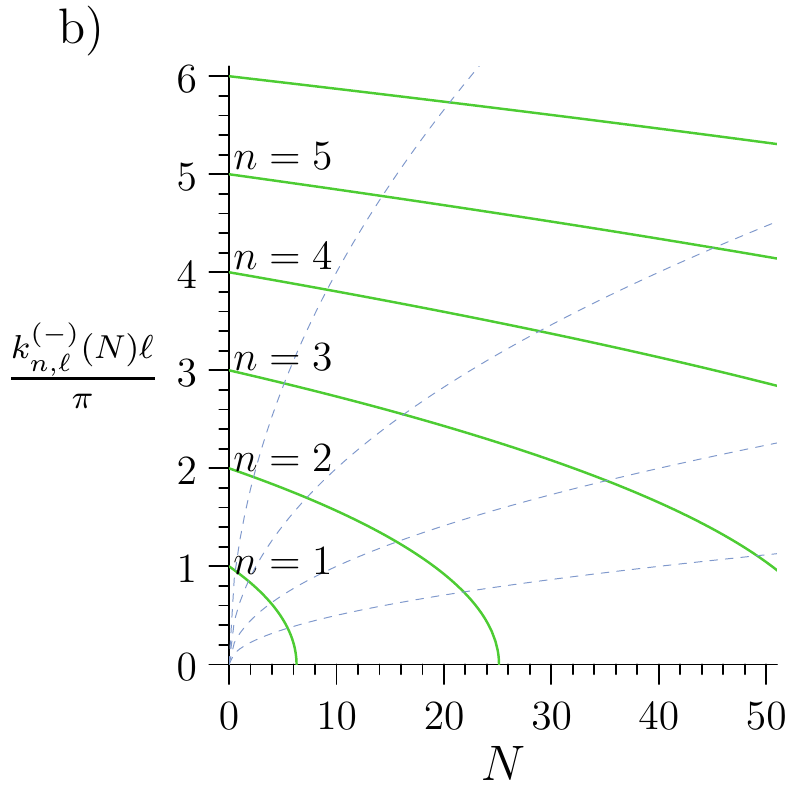}
\caption{ \label{fig3}
  The spectral curves $k^{(\pm)}_{n,\ell}(N)$ (full green lines)
  in the repulsive (\textbf{a}) and
  attractive case (\textbf{b}).
  The~$n$-th curve is obtained from the curve for $n=1$ by scaling
  $k^{(\pm)}_{n,\ell}(N)= n k^{(\pm)}_{1,\ell}(N/n^2)$.
  The (blue) dashed lines indicate trajectories of the flow
  \eqref{eq:flow}.
  The deformation parameter $m$ is constant along
  the~flow.
}
\end{figure}

{By direct calculation (see \cite{GW2}) we express the $L^{2}$-norms
in terms of elliptic integrals (see~Appendix~\ref{sec:elliptic})~as}
\begin{equation}
  \begin{split}
    N^{(+)}_{n,\ell}(m)=&
    \frac{8n^{2}}{\ell}\ K(m)\left[K(m)-E(1,m)\right]
    \\[0.3cm]
    N^{(-)}_{n,\ell}(m)=&
    \frac{8n^{2}(1-m)}{\ell}\ K(m)\left[\Pi(1,m,m)-K(m)\right],
  \end{split}
  \label{N_m}
\end{equation}
and use those to implicitly define the spectral curves
in the form $k^{(\pm)}_{n,\ell}(N)$.
The latter spectral curves are shown in Figure~\ref{fig3}.
The monotonicity of the spectral curves in this form follows
from the monotonicity of $k^{(\pm)}_{n,\ell}(m)$ together with the
monotonicity of $N^{(\pm)}_{n,\ell}(m)$.
More precisely, one may check that $N^{(\pm)}_{n,\ell}(m)$ in (\ref{N_m})
is an increasing function of $m$ in the corresponding interval $m\in(0,1]$
for $g=1$ and $m\in\left(0,\frac{1}{2}\right)$ for $g=-1$.
To verify this statement, observe that
\begin{enumerate}
\item $\left[K(m)-E(1,m)\right]$ and
  $\frac{1}{m}\left[\Pi(1,m,m)-K(m)\right]$
  are increasing functions of $m$. This follows from their integral
  representations (see Appendix~\ref{sec:elliptic}).
  Explicitly, writing each expression as an integral,
  the corresponding integrands are positive and pointwise increasing
  functions of $m$.
\item $K(m)$ and $m(1-m)$ are also positive increasing functions of $m$ in the relevant intervals.
\end{enumerate}

The inverse of $N^{(\pm)}_{n,\ell}(m)$
will be denoted $m^{(\pm)}_{n,\ell}(N)$. Combining the monotonicity of
$k^{(\pm)}_{n,\ell}(m)$ and $N^{(\pm)}_{n,\ell}(m)$ one finds
 in the
repulsive case that $k^{(+)}_{n,\ell}(N)$
is an increasing function of $N$ defined for $N>0$
while
in the
attractive case $k^{(-)}_{n,\ell}(N)$
is a decreasing function defined on $0<N<N_{n,\ell}^{(-),\mathrm{max}}$
where
\begin{equation}
  N_{n,\ell}^{(-),\mathrm{max}}=
  \frac{4n^{2}}{\ell}K\left(\frac{1}{2}\right)\left[\Pi\left(1,\frac{1}{2},\frac{1}{2}\right)-K\left(\frac{1}{2}\right)\right]\ .
\end{equation}

A characterization of the spectral curves $k^{(\pm)}_{n,\ell}(N)$ may be given as follows. We define the following flow in the $k$-$N$-plane (see Figure \ref{fig3}).
\begin{equation}
  \Phi^\tau(m)=(N^{(\pm)}_{\tau,\ell}(m), k^{(\pm)}_{\tau,\ell}(m)),
\label{eq:flow}
\end{equation}
where $N^{(\pm)}_{\tau,\ell}(m)$ and $k^{(\pm)}_{\tau,\ell}(m)$ are
extensions
of the expressions in Equations (\ref{N_m}) and (\ref{k_m}), replacing the
integer valued $n$ with the real flow parameter $\tau$. Observe that
$k^{(\pm)}_{\tau,\ell}(m)$ depends linearly on $\tau$ whereas,
$N^{(\pm)}_{\tau,\ell}(m)$
is proportional to $\tau^2$.
This means that for each value of $m$, the corresponding flow line
$\left\{\Phi^\tau(m)\right\}_{\tau=0}^{\infty}$ is of the form
$k=\gamma\sqrt{N}$
(where $\gamma$ depends on $m$).
In particular, this implies that the spectral curves $k^{(\pm)}_{n,\ell}(N)$ are self-similar
\begin{equation}
  k^{(\pm)}_{n,\ell}(N)=nk^{(\pm)}_{1,\ell}\left(\frac{N}{n^{2}}\right).
\end{equation}

In addition, each flow line traverses the spectral curves $k^{(\pm)}_{n,\ell}(N)$
in the order given by the number of nodal domains $n$. This
implies that the spectral curves never cross each other and remain properly~ordered. We thus obtain the following second
generalization of Sturm's oscillation theorem on the~interval.

\begin{Proposition} For $g=1$ (repulsive case) let $N>0$ and for $g=-1$
(attractive case) let $N\in(0,N_{1,\ell}^{(-),\mathrm{max}})$.\\
Then there is a discrete set $\{k_{n}\}_{n=1}^{\infty}$ of positive
real numbers, increasingly ordered such that
\vspace{6pt}
\begin{equation*}
  \phi_{n}=\left. \chi^{(\pm)}_{m^{(\pm)}_{n,\ell}(N),\,k_n}\right|_{[0,\ell]}
\end{equation*}
{is a solution of the NLS equation on the interval with a spectral parameter $\mu_{n}=k_{n}^{2}$ and
$L^{2}$-norm \mbox{$N=\int_{0}^{\ell}\phi_{n}(x)^{2}\ dx$} and $n$ is the number of nodal domains.
Furthermore, these are all solutions whose $L^{2}$-norm equals $N$.}

\end{Proposition}

\subsection{Nonlinear Quantum Star Graphs}

\begin{figure}[h]
\centering
  \includegraphics[width=0.495\textwidth]{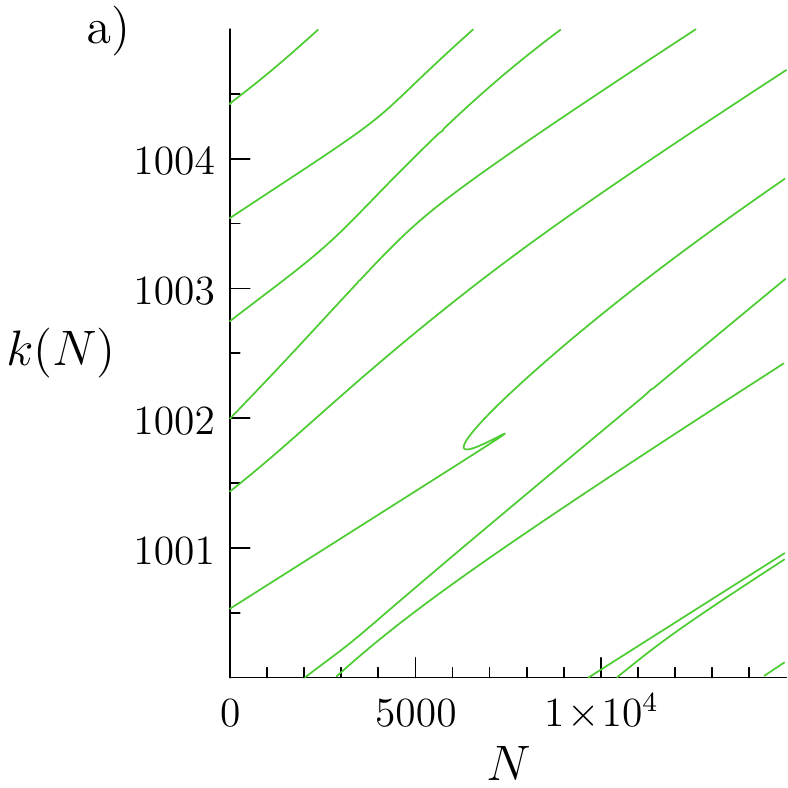}
  \includegraphics[width=0.495\textwidth]{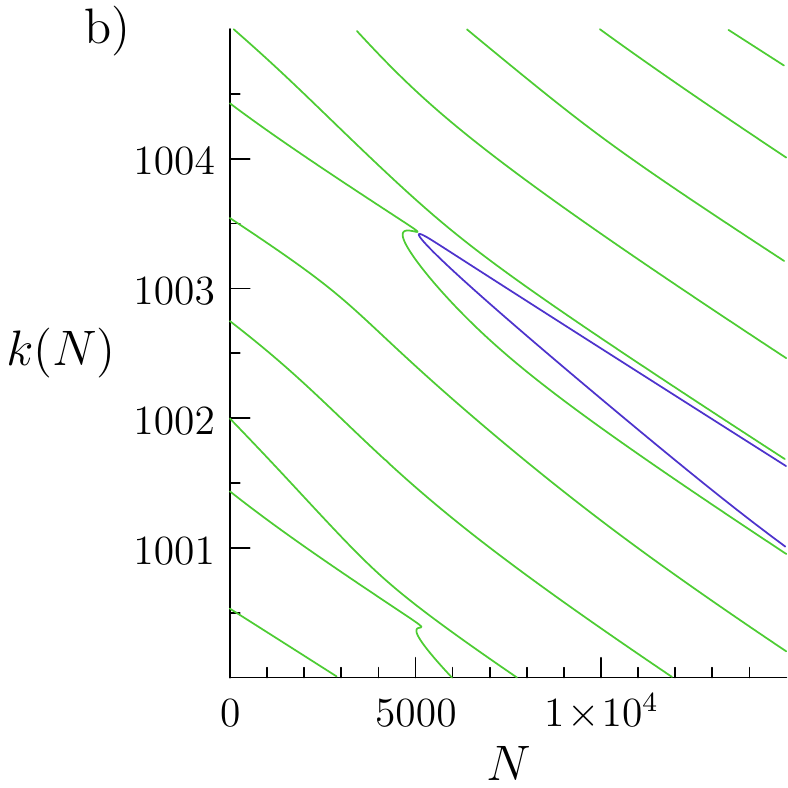}
  \caption{\label{fig4}
    Spectral curves $k(N)$ for a nonlinear star
    graph with $E=3$ edges and edge lengths $\ell_1=1$,
    $\ell_2=\sqrt{2}$, and $\ell_3=\sqrt{3}$ with repulsive (\textbf{a}) and
    attractive (\textbf{b}) nonlinear interaction. The spectral curves have
    been obtained
    by numerically solving the matching conditions
    using a Newton-Raphson method. For $N \to 0$ one obtains the
    spectrum
    of the corresponding linear star graph. Apart from one curve, all
    shown curves are connected to the linear spectrum this way. In the
    attractive case one spectral curve (shown in blue) is not
    connected to the linear spectrum. Such curves can sometimes be
    found by coincidence, e.g., if one is close to a bifurcation and
    numerical
    inaccuracy allows to jump from one solution branch to another
    (and this is indeed how we found it).
    In the repulsive case there is one spectral curve that has a sharp
    cusp. This indicates that there may be a bifurcation nearby that
    has additional solution branches that have not been found. In
    general it is a non-trivial numerical task to ensure that a
    diagram of spectral curves is complete. Here, completeness has not
    been attempted as the picture serves a mainly illustrative purpose.
  }
\end{figure}

One may use the functions $\chi^{(\pm)}_{m,k}(x)$ defined in Equation~\eqref{chi_sol}
in order to reduce the problem of finding a solution of the NLS equation
on a star graph to a (nonlinear) algebraic problem. By setting
\vspace{6pt}
\begin{equation}
  \phi_{e}(x_{e})=\sigma_{e}\chi^{(\pm)}_{m_{e},k}(\ell_{e}-x_{e})
\end{equation}
where an overall sign $\sigma_{e}=\pm 1$ and the deformation parameter
$m_{e}$ remain unspecified (and allowed to take different values
on different edges) one has a set of $E$ functions that satisfy
the NLS equation with spectral parameter $\mu=k^{2}$ on each edge
and also satisfy the Dirichlet condition $\phi_{e}(\ell_{e})=0$ at the boundary vertices.
Setting $\sigma_{e}=\mathrm{sgn}\left(\chi^{(\pm)}_{m_{e},k}(\ell_{e})\right)$
(unless $\chi^{(\pm)}_{m_{e},k}(\ell_{e})=0$)
the Kirchhoff matching conditions at the centre give a set of $E$ independent nonlinear algebraic equations
(see Equations~\eqref{eq:Kirchhoff-1} and \eqref{eq:Kirchhoff-2})
for $E$ continuous parameters $\{m_{e}\}$. If $k$ is fixed there
are typically discrete solutions for the parameters $\{m_{e}\}$.
As $k$ varies the solutions deform and form one-parameter families.
Setting
\begin{equation}
  N=\sum_{e=1}^{E}\int_{0}^{\ell_{e}}\phi_{e}(x_{e})^{2}\ dx_{e}
\end{equation}
each solution may be characterized by a
pair $(k,N)$ and as $k$ is varied one naturally arrives at
spectral curves in the $k$-$N$-plane, that may be expressed as $k(N)$ (or $N(k)$), as we have seen
for the interval in the previous section. Nevertheless,
the spectral curves of the star graph have a more intricate structure (see Figure \ref{fig4}).
In non-linear algebraic equations one generally expects that solutions
appear or disappear in bifurcations. For any particular example some numerical approach is needed to find the spectral curves.
To do so, one first needs to have some approximate solution (either found by analytical
approximation or by a numerical search in the parameter space). After that Newton-Raphson methods may be used to find the solution up to the desired numerical accuracy and the spectral curves are found by varying the spectral parameter slowly.

Figure~\ref{fig4} shows spectral curves that have been found numerically
for a star graph with $E=3$ and edge lengths $\ell_{e}=\sqrt{e}$
($e=1,2,3$). Most of the curves have been found starting from the
corresponding spectrum of the linear problem ($g=0$).
Yet, one can see an additional curve that does
not connect to the linear spectrum as $N\to 0$.
This has originally been found
in previous work \cite{GW2} by coincidence, as~the
the numerical method jumped from one curve to another where they almost touch
in the diagram.

{We stress that in a numerical approach it is very hard to make sure that
all solutions of interest are~found, even if one restricts the
search to a restricted region in parameter space.
A full characterization
of all solutions (such as given above for the nonlinear interval)
will generally be elusive even for basic nonlinear quantum graphs.
Theorems~\ref{thm1} and \ref{thm2} and the
related Corollaries \ref{cor1} and~\ref{cor2} establish the existence
of a large set of solutions inside the deep nonlinear regime. Each
of these solutions may be used as a~starting point for a numerical
calculation of further solutions along the corresponding spectral
curves.
}

\subsection{Nodal Edge Counting and Central Dirichlet Solutions}

\begin{figure}[h]
\centering
   \includegraphics[width=0.75\textwidth]{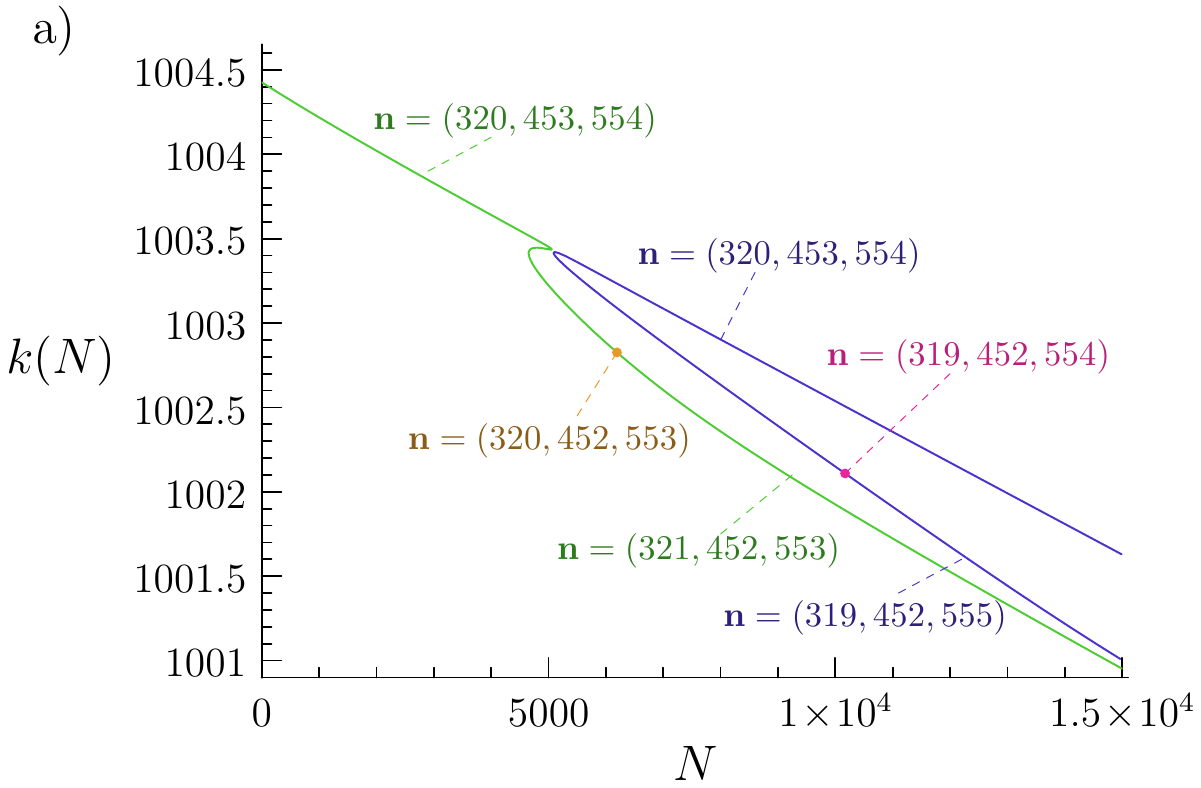}\\[0.3cm]
   {\large  $\left. b\right)$ }\hspace*{15cm}\hfill\\
     \includegraphics[width=0.32\textwidth]{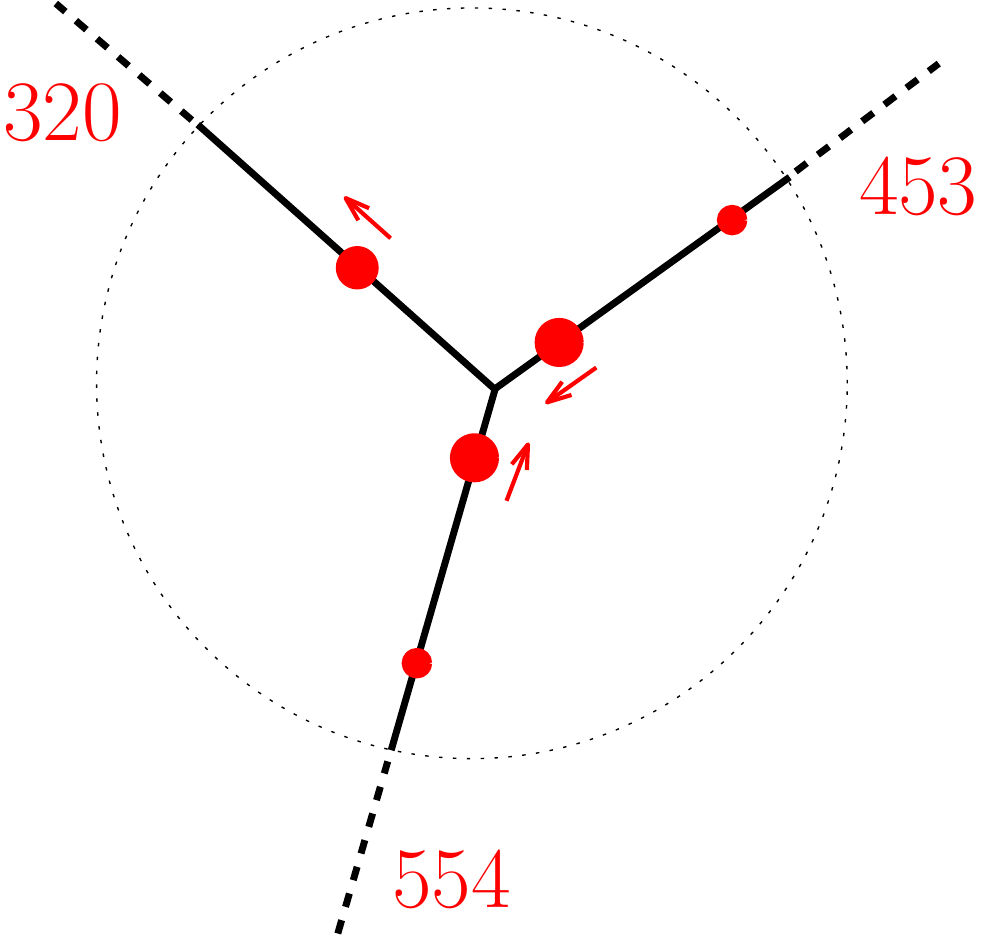}
   \hfill
   \includegraphics[width=0.32\textwidth]{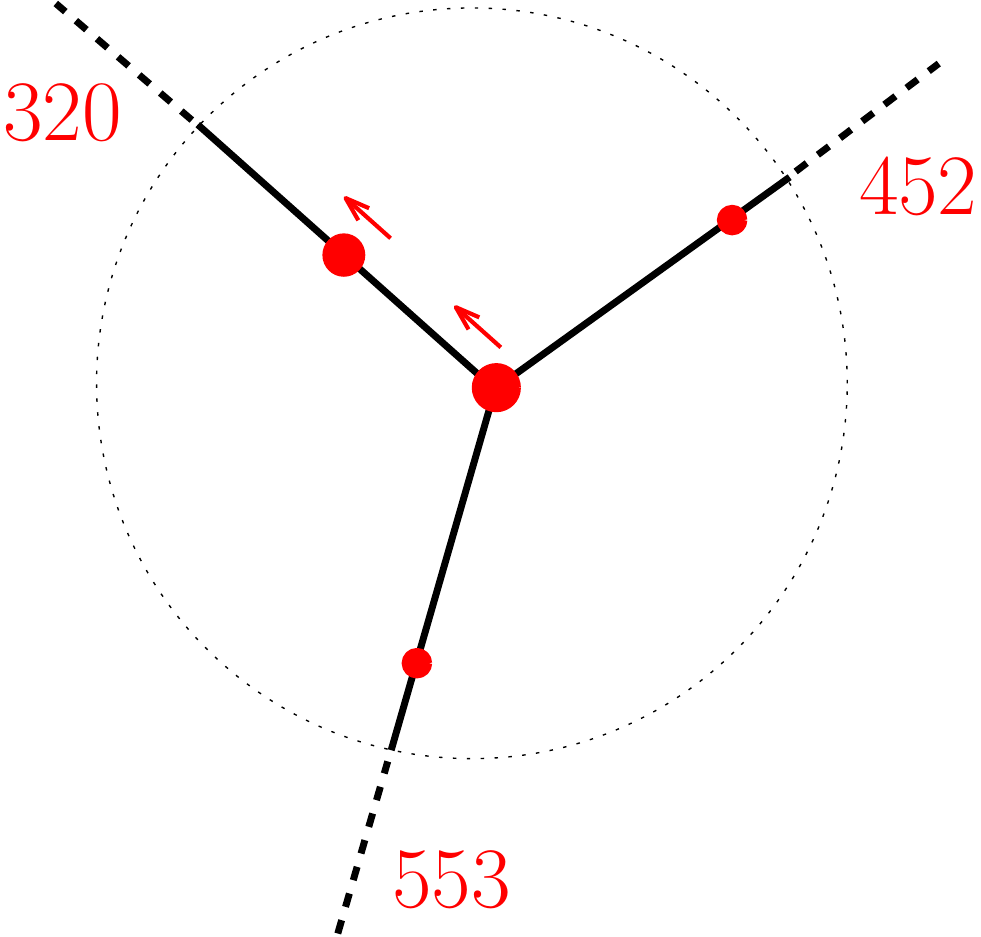}
   \hfill
   \includegraphics[width=0.32\textwidth]{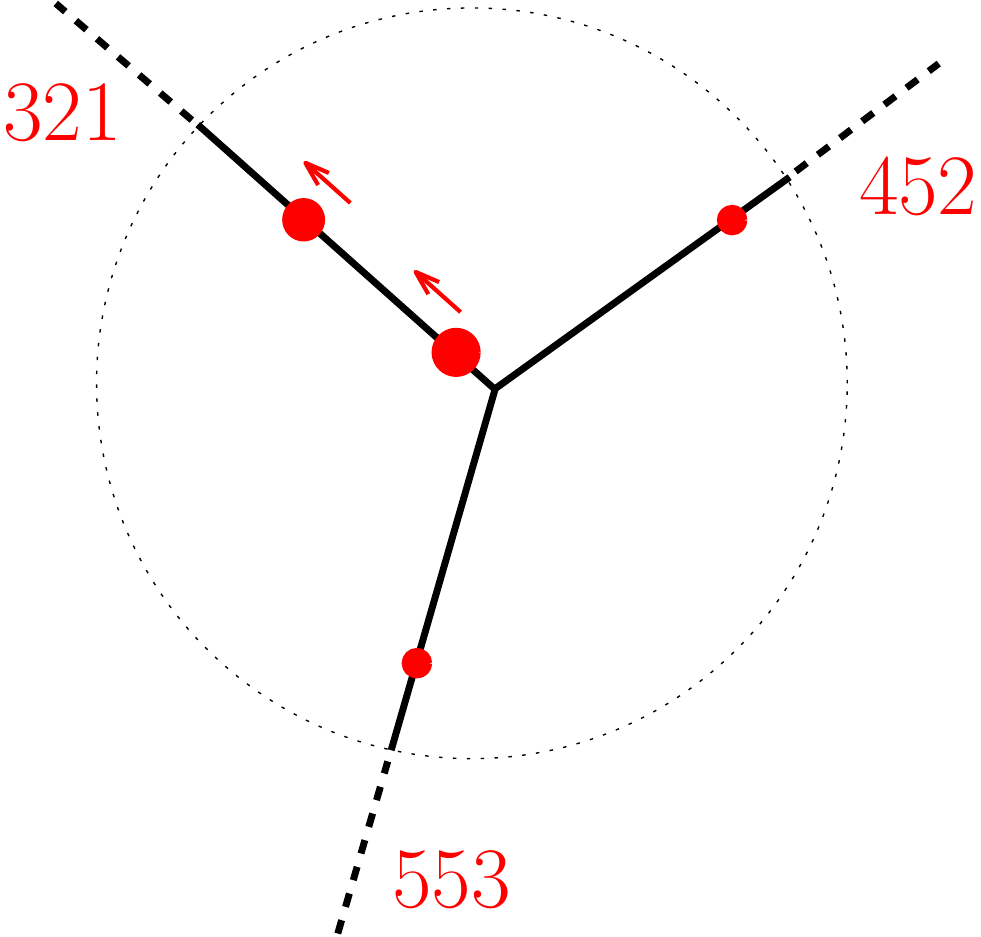}
\caption{
  \label{fig5}
  Upper panel (\textbf{a}):
  Two spectral curves (green and blue) of the star
  graph
  described in the
  caption of Figure~\ref{fig4} with attractive nonlinear
  interaction. The yellow and pink dots indicate positions that
  correspond to central
  Dirichlet solutions. The nodal edge structure $\mathbf{n}$ is
  indicated for each part of the curve. The latter is
  constant along spectral curves apart from jumps at the positions
  that correspond to central Dirichlet solutions. 
  Lower panel (\textbf{b}): The three diagrams show how the nodal points move through the
  centre while $N$ is increased through a central Dirichlet point on a
  spectral curve (the green curve in the upper panel).
  Only some nodal points close to the centre
  are shown. In the left diagram the three  large dots are the closest to the centre
  and the arrows indicate how they move when $N$ is increased.
  The numbers give the number of nodal domains
  on each edge.
  Increasing $N$ further two nodal points on different edges merge at
  the centre as shown in the middle diagram. On the corresponding
  edges one nodal domain disappears.
  Further increasing $N$ the nodal point moves from the centre into
  the remaining edge where the number of nodal domains is increased by
  one.
}
\end{figure}

{It is interesting to consider the nodal structure along a spectral
curve. Generically the wavefunction does not vanish at the centre
and the nodal edge count structure (i.e., the vector $\mathbf{n}$)
remains constant along the curve.
The existence of central Dirichlet solutions implies that nodal points
may move into (and through) the centre along a spectral curve
(see also Theorem 2.9, \cite[]{BanKru_amspro18}). At this instance the nodal
edge count structure changes twice; first when the node hits the centre
and then again when it has moved through. If $\mathbf{n}_{0}$ is
the nodal edge count structure at a central Dirichlet solution, then
generically the value of the function at the centre will change
its sign along the spectral curve close to the central Dirichlet solution.
If $\mathbf{n}_{<}$ and $\mathbf{n}_{>}$ are the nodal edge count
structures close to the central Dirichlet solution then their entries
differ at most by one $n_{>,e}-n_{<,e}=\pm1$ and when the nodal point
hits the centre one has $n_{o,e}=\mathrm{min}(n_{>,e},n_{<,e})$.
This is shown in more detail for a numerical example in Figure~\ref{fig5},
where some central Dirichlet solutions are indicated on the spectral curves.
The figure also shows the relevance of
the central Dirichlet solutions for finding numerical solutions. The
central Dirichlet solutions can be constructed directly using the
machinery of the proof in the next section. From that one can then
obtain a full spectral curve numerically by varying the parameters
appropriately.}

\section{Proofs of Main Theorems}

\label{sec:proofs}

{We  prove the two theorems for repulsive and attractive interaction separately.~The main construction is however the same. We start by describing the idea behind the construction and then turn to the actual proofs.
Let $m^{(\pm)}_{1,\ell_{e}}(k)$ be the functions describing the
deformation parameter
of solutions on the interval of length $\ell$ with Dirichlet boundary
conditions and a
single nodal domain (they are given as the inverse of Equation~(\ref{k_m}); see
also the
lowest curve in Figure \ref{fig2}). Those functions are well-defined
for
$k>\frac{\pi}{\ell_{e}}$
in the repulsive case and for $k<\frac{\pi}{\ell_{e}}$ in the
attractive case.
Using these, we~define}
\vspace{6pt}
\begin{equation}
  \phi_{e}:=\sigma_{e} \left.\chi^{(\pm)}_{m^{(\pm)}_{1,\ell_{e}}(k),\,k}\right|_{[0,\ell_e]},
\end{equation}
for $e=1,\dots,E$, and where $\sigma_{e}=\pm 1$ are signs that will
be specified later. These are of course just the solutions
of the NLS with one nodal domain on the corresponding interval at
spectral parameter $\mu=k^{2}$.
In order to ensure that this function
is well defined on all edges at given $k$ we have to
choose $k\in (\frac{\pi}{\ell_{1}}, \infty)$
in the repulsive case and $k\in(0, \frac{\pi}{\ell_{E}})$ in the
attractive case
(recall that we ordered the edge lengths by $\ell_{1} < \ldots < \ell_{E}$).

As $\phi_{e}(0)=0$ by construction, the set $\left\{\phi_e\right\}_{e=1}^{E}$ defines a continuous function
on the graph including the centre for all allowed values of $k$.
However, in general, these functions
do not satisfy the remaining Kirchhoff condition
$\sum_{e=1}^{E}\frac{d\phi_{e}}{dx_{e}}(0)=0$.
The idea of the proofs is the following.
We consider $\sum_{e=1}^{E}\frac{d\phi_{e}}{dx_{e}}(0)$ as a function
of $k$
and need to show that it vanishes at some $k=k_0$.
We find a particular set of signs $\{\sigma_{e}\}$ for which it is
easy to show that
$\sum_{e=1}^{E}\frac{d\phi_{e}}{dx_{e}}(0)$ changes sign as $k$ is
varied
within its allowed range.
Since this function is continuous in $k$ it must vanish somewhere,
which establishes the required central Dirichlet solution with exactly one nodal domain on each edge.

As we recognized above the role which the derivative of the solution
plays in the proof, let us now directly calculate it.
\vspace{6pt}
\begin{equation}
  \theta^{(\pm)}(m):=\frac{1}{\sqrt{2}k^{2}}\frac{d\chi^{(\pm)}_{m,k}}{dx}(0)=
  \begin{cases}
    \frac{\sqrt{m}}{1+m} & \text{for \ensuremath{g=1},}\\[0.3cm]
    \frac{\sqrt{m(1-m)}}{1-2m} & \text{for \ensuremath{g=-1}.}
  \end{cases}\label{theta_g}
\end{equation}

In particular,
expressing the derivative as a function of $m$ and $k$ and
multiplying by a factor  $\frac{1}{\sqrt{2}k^{2}}$,
we see that the resulting function $\theta^{(\pm)}(m)$
does not depend explicitly on $k$, but only via the deformation
parameter, $m$.


\begin{Remark} \label{rem: Equal-edge-lengths}

In the statement of the theorem we have assumed that all edge lengths
are different and stated how this may be relaxed in a subsequent remark.
We can explain this now in more detail. Assume that we have two edge lengths
that coincide.
Denote those edges by $e_{0},e_{1}$ and follow the construction
above. By choosing opposite sign for the two edges
$\sigma_{e_{0}}=\-\sigma_{e_{1}}$
the contribution of the two edges to the sum of derivatives
$\sum_{e=1}^{E}\frac{d\phi_{e}}{dx_{e}}(0)$
cancels exactly for all allowed values of $k$,
that is
$\frac{d\phi_{e_{0}}}{dx_{e_{0}}}(0)+\frac{d\phi_{e_{1}}}{dx_{e_{1}}}(0)=0$.
One may then focus on the subgraph where the two edges are deleted
and continue to construct a solution on the subgraph.\\
 One may also start with a graph with different edge lengths. If one
has found any regular central Dirichlet solution on the graph one
may add as many pairs of edges of the same length and find a regular
non-Dirichlet solution on the larger graph following the above construction.
\end{Remark}

\subsection{The Repulsive Case $g=1$}

\label{sec:proof_thm1}

\begin{proof}[{Proof of Theorem~\ref{thm1}}]
Using the construction defined above we have to establish that there
is a choice for the signs $\boldsymbol{\sigma}=(\sigma_{1},\dots,\sigma_{E})$
and value for the spectral parameter $k=\sqrt{\mu}$ such that the Kirchhoff
condition $\sum_{e=1}^{\infty}\frac{d\phi_{e}}{dx_{e}}(0)=0$ is satisfied.
For $k\in(\frac{\pi}{\ell_{1}},\infty)$ let
us define the function
\begin{equation}
  f_{\boldsymbol{\sigma}}(k):=\sum_{e=1}^{E}\sigma_{e}\theta^{(+)}
  \left(m^{(+)}_{1,\ell_{e}}(k)\right)=
  \sum_{e}^{E} \sigma_e \frac{\sqrt{m^{(+)}_{1,\ell_{e}}(k)}}{1+m^{(+)}_{1,\ell_{e}(k)}}
\end{equation}
where $\theta^{(+)}(m)$ was defined in Equation~(\ref{theta_g}) and $m^{(+)}_{1,\ell_{e}}(k)$
is the inverse of
\begin{equation}
  k^{(+)}_{1,\ell_{e}}(m)=\frac{2}{\ell_{e}}\sqrt{1+m}K(m)
  \label{eq:k_plus_m}
\end{equation}
as defined in Equation~\eqref{k_m} (setting $n=1$ for one nodal domain). The
Kirchhoff condition is equivalent to the condition $f_{\boldsymbol{\sigma}}(k)=0$
for some $k>\frac{\pi}{\ell_{1}}$.

{To continue the proof, we point out some monotonicity properties of $\theta^{(+)}(m)$ and
$m^{(+)}_{1,\ell_{e}}(k)$. These~properties may be easily verified by direct calculation using Equations~(\ref{k_m}) and (\ref{theta_g}).
For~$m\in(0,1)$ the functions $\theta^{(+)}(m)$ and $k^{(+)}_{1,\ell_{e}}(m)$
are strictly increasing and}
\begin{align}
\theta^{(+)}(0)=0, & \quad \theta^{(+)}(1)=\frac{1}{2}, \\
k^{(+)}_{1,\ell_{e}}(0)=\frac{\pi}{\ell_{e}}, & \quad
k^{(+)}_{1,\ell_{e}}(m)\underset{m\to 1}{\longrightarrow}\infty.
\end{align}

This implies that $\theta^{(+)}\left(m^{(+)}_{1,\ell_{e}}(k)\right)$
is strictly increasing for $k \in(\frac{\pi}{\ell_{e}}, \infty)$ and
\begin{align}
\theta^{(+)}\left(m^{(+)}_{1,\ell_{e}}\left(\frac{\pi}{\ell_{e}}\right)\right)=0
\quad , \quad \theta^{(+)}\left(m^{(+)}_{1,\ell_{e}}(k)\right) \underset{k\to\infty}{\longrightarrow}\frac{1}{2}.
\label{eq:theta_plus_properties}
\end{align}

If $E\ge 3$ is odd we choose the signs $\boldsymbol{\sigma}=(\sigma_{1},\dots,\sigma_{E})$ to satisfy
the following conditions
\vspace{6pt}
\begin{equation}
\sigma_{1}=-1\quad \textrm{and} \quad \sum_{e=2}^{E}\sigma_{e}=0
\label{eq:sign_choice} \end{equation}
and
\begin{equation}
  f_{\boldsymbol{\sigma}}\left(\frac{\pi}{\ell_{1}}\right)=\sum_{e=1}^{E}\sigma_{e}\theta^{(+)}\left(m^{(+)}_{1,\ell_{e}}\left(\frac{\pi}{\ell_{1}}\right)\right)>0\ .
\end{equation}

Such a choice of signs is always possible as $\theta^{(+)}\left(m^{(+)}_{1,\ell_{e}}\left(\frac{\pi}{\ell_{1}}\right)\right)>0$
for all $e>1$.
We then get by Equations~(\ref{eq:theta_plus_properties}) and (\ref{eq:sign_choice}) that  
$\lim_{k\to\infty}f_{\boldsymbol{\sigma}}(k)=\frac{1}{2}\sum_{e=1}^{E}\sigma_{e}=-\frac{1}{2}$.
By continuity there exists $k_{0}\in(\frac{\pi}{\ell_{1}}, \infty)$ such that $f_{\boldsymbol{\sigma}}(k_{0})=0$
for the given choice of signs. This proves the theorem for odd $E$. 

{For an even number
of edges $E=2M$ ($M\ge2$) one needs to do a little bit more work.
In this case, there are two strategies for choosing signs, $\boldsymbol{\sigma}=(\sigma_{1},\dots,\sigma_{E})$,
and showing that $f_{\boldsymbol{\sigma}}\left(k\right)$ vanishes for some $k$.}
\begin{enumerate}
\item
  One may choose more negative signs than positive signs so
  that $\sum_{e}\sigma_{e}<0$.
  Then $\lim_{k\to\infty}
  f_{\boldsymbol{\sigma}}\left(k\right)=\frac{1}{2}\sum_{e=1}^{E}\sigma_{e}$
  is trivially negative.
  The difficulty here is in showing that such a choice is consistent with
  $f_{\boldsymbol{\sigma}}\left(\frac{\pi}{\ell_{1}}\right)>0$.
  This generally leads to some conditions which the edge lengths should satisfy.
\item One may choose as many positive as negative signs, which makes it easier to satisfy $f_{\boldsymbol{\sigma}}\left(\frac{\pi}{\ell_{1}}\right)>0$ (i.e., the conditions on the edge lengths are less restrictive).
  Yet, the difficulty here lies in $\lim_{k\to\infty}f_{\boldsymbol{\sigma}}\left(k\right)=0$,
  which means that one needs to show that this limit is approached from the
  negative side (i.e., find the conditions on the edge lengths which ensures this).
\end{enumerate}

These two strategies give some indication on how our proof may be generalized beyond
the stated length restrictions. Moreover, they also give a practical instruction
for how one may search for further solutions numerically.

We continue the proof by following the second strategy
and setting
\begin{equation}
  \sigma_{e}=\begin{cases}
    1 & \text{for \ensuremath{e=1} and \ensuremath{e\ge M+2},}\\
    -1 & \text{for \ensuremath{2\le e\le M+1}}
  \end{cases}
  \label{sign_choice_even_edges}
\end{equation}
so that $\sum_{e=1}^{E}\sigma_{e}=0$. One then has $\lim_{k\to\infty}f_{\boldsymbol{\sigma}}\left(k\right)=0$
and we will show that the leading term in the
(convergent) asymptotic expansion of $f_{\boldsymbol{\sigma}}\left(k\right)$ for large
$k$ is negative. Using the known asymptotics \cite{NIST} of the elliptic
integral $K(m)$ as $m=1-\delta m$ goes to one (or $\delta m\to 0$)
\vspace{6pt}
\begin{equation}
  K(1-\delta m)=-\frac{1}{2}\log(\delta m)+2\log(2)+O\left(\delta m\ \log(\delta m)\right)
\end{equation}
one may invert Equation~\eqref{eq:k_plus_m} asymptotically for large $k$ as
\begin{equation}
  1-m^{(\pm)}_{1,\ell_e}(k)=16e^{-\frac{k\ell_{e}}{\sqrt{2}}}+O\left(ke^{-\sqrt{2}k\ell_{e}}\right)
\end{equation}
and,   thus
  \begin{equation}
    \begin{split}
      \frac{\sqrt{m^{(\pm)}_{1,\ell_e}(k)}}{1+m^{(\pm)}_{1,\ell_e}(k)}
      =&
      \frac{\sqrt{1-\left(1-m^{(\pm)}_{1,\ell_e}(k)\right)}}{
        2-\left(1-m^{(\pm)}_{1,\ell_e}(k)\right)}\\
      =&\frac{1}{2}-\frac{1}{16}
      \left(1-m^{(\pm)}_{1,\ell_e}(k)\right)^2 +
      O\left( \left(1-m^{(\pm)}_{1,\ell_e}(k)\right)^3\right)\\
      =&
      \frac{1}{2} - 16e^{-\sqrt{2} k\ell_{e}}+
      O\left(ke^{-3\frac{k\ell_{e}}{\sqrt{2}}}\right)\ .
    \end{split}
  \end{equation}

This directly leads to the asymptotic expansion
\vspace{6pt}
\begin{small}
  \begin{equation}
    f_{\boldsymbol{\sigma}}(k)=
    -\sum_{e=1}^{E}\sigma_{e}16e^{-\sqrt{2} k\ell_{e}}+
    O\left(ke^{-3\frac{k\ell_{1}}{\sqrt{2}}}\right)
    =-16e^{-\sqrt{2} k\ell_{1}}
    \left(1+\sum_{e=2}^{E}\sigma_{e}
      e^{-\sqrt{2}k(\ell_{e}-\ell_{1})}\right)
    +O\left(ke^{-3 \frac{k\ell_{1}}{\sqrt{2}}}\right)
  \end{equation}
  \end{small}
which is negative for sufficiently large $k$ because $\ell_{1}$
is the shortest edge length.

It is left to show
$f_{\boldsymbol{\sigma}}\left(\frac{\pi}{\ell_{1}}\right)>0$.

  For this let us write
  $m^{(+)}_{1,\ell_e}(k)=m^{(+)}_{1,1}(k\ell_e)$ for each term.
  As $\theta^{(+)}\left(m^{(+)}_{1,1}\left(\pi\right)\right)=0$
  the condition $f_{\boldsymbol{\sigma}}\left(\frac{\pi}{\ell_{1}}\right)>0$
  is equivalent to
\begin{equation}
  \sum_{e=2}^{M+1}\theta^{(+)}
  \left(m^{(+)}_{1,1}\left(\frac{\pi\ell_{e}}{\ell_{1}}\right)\right)
  <
  \sum_{e=M+2}^{E}\theta^{(+)}
  \left(m^{(+)}_{1,1}\left(\frac{\pi\ell_{e}}{\ell_{1}}\right)\right),
  \label{cond1}
\end{equation}
using our choice of the signs, Equation~(\ref{sign_choice_even_edges}).
Since $m^{(+)}_{1,1}$ and
$\theta^{(+)}$
are increasing functions and \mbox{$\ell_1<\ldots<\ell_E$}, condition \eqref{cond1}
is certainly satisfied if
\begin{equation}
  \frac{\theta^{(+)}\left(m^{(+)}_{1,1}
      \left(\frac{\pi\ell_{M+2}}{\ell_{1}}\right)\right)}{
    \theta^{(+)}\left(m^{(+)}_{1,1}
      \left(\frac{\pi\ell_{M+1}}{\ell_{1}}\right)\right)}>\frac{M}{M-1}\ .
  \label{cond2}
\end{equation}

The condition \eqref{cond2} restricts the three edge lengths $\ell_{1}$, $\ell_{M+1}$ and
$\ell_{M+2}$ and it is equivalent to the condition \eqref{cond_thm1}
stated in the theorem.
Indeed, this is trivial for the right-hand side where \mbox{$\frac{M}{M-1}=\frac{E}{E-2}$}. For~the left-hand side
  note that Equation~\eqref{eq:thm1_mplusminus} in Theorem~\ref{thm1} identifies
  $m_+= m^{(+)}_{1,1}
  \left(\frac{\pi\ell_{M+2}}{\ell_{1}}\right)$ and \mbox{$m_-=m^{(+)}_{1,1}
  \left(\frac{\pi\ell_{M+1}}{\ell_{1}}\right)$} such that the left-hand-side of
  the stated condition \eqref{cond_thm1} in the theorem and the left-hand side of
  Equation~\eqref{cond2} are identical when written out explicitly.
\end{proof}

\subsection{The Attractive
Case $g=-1$}

\label{sec:proof_thm2}

\begin{proof}[ {Proof of Theorem~\ref{thm2}}]
In the attractive case we can start similarly to the previous proof by rewriting
the Kirchhoff condition on the sum of
derivatives as $f_{\boldsymbol{\sigma}}(k)=0$
for some $k\in(0,\frac{\pi}{\ell_{E}})$ where
\begin{equation}
  f_{\boldsymbol{\sigma}}(k)=
  k^{2}
  \sum_{e=1}^{E}\sigma_{e}\theta^{(-)}\left(m^{(-)}_{1,\ell_{e}}(k)\right)
  =\sum_{e}^{E}\sigma_{e}\frac{4\sqrt{m^{(-)}_{1,\ell_{e}}(k)
      (1-m^{(-)}_{1,\ell_{e}}(k))}
    K\left( m^{(-)}_{1,\ell_{e}}(k)  \right)^{2}}{\ell_{e}^{2}}\ .
\end{equation}

The additional factor $k^{2}$ is irrelevant for satisfying the condition
but allows us to extend the definition of the function to $k=0$ (where
$k^{2}\sim1-2m^{(-)}_{1,\ell}(k)$).
Noting that $m^{(-)}_{1,\ell_{e}}(k)$ is a
decreasing function for
$k\in(0,\frac{\pi}{\ell_{e}})$
with $m^{(-)}_{1,\ell_{e}}(0)=\frac{1}{2}$
and $m^{(-)}_{1,\ell_{e}}\left(\frac{\pi}{\ell_{e}}\right)=0$ and $K(m)$ is increasing with $m$ we get that
the function
\begin{equation*}
k^{2}\theta^{(-)}\left(m^{(-)}_{1,\ell_{e}}(k)\right)
=\frac{4\sqrt{m^{(-)}_{1,\ell_{e}}(k)(1-m^{(-)}_{1,\ell_{e}}(k))}
  K\left( m^{(-)}_{1,\ell_{e}}(k)  \right)^{2}}{\ell_{e}^{2}}
\end{equation*}
is a decreasing function for $k\in(0,\frac{\pi}{\ell_{e}})$ and
\begin{equation*}
\lim_{k\to 0}k^{2}\theta^{(-)}\left(m^{(-)}_{1,\ell_{e}}(k)\right)=\frac{2K(\frac{1}{2})^{2}}{\ell_{2}^{2}}
\quad , \quad
\left(\frac{\pi}{\ell_{e}}\right)^{2}\theta^{(-)}\left(m^{(-)}_{1,\ell_{e}}\left(\frac{\pi}{\ell_{e}}\right)\right)=0.
\end{equation*}

Altogether this implies that
\begin{equation}
  f_{\boldsymbol{\sigma}}(0)
  =2K\left(\frac{1}{2}\right)^{2}\sum_{e=1}^{E}\frac{\sigma_{e}}{\ell_{e}^{2}}
\end{equation}
and
\begin{equation}
  f_{\boldsymbol{\sigma}}\left(\frac{\pi}{\ell_{E}}\right)
  =\sum_{e=1}^{E-1}\sigma_{e}
  \frac{4\sqrt{m^{(-)}_{1,\ell_e}\left(\frac{\pi}{\ell_{E}}\right)
      \left(1-m^{(-)}_{1,\ell_e}\left(\frac{\pi}{\ell_{E}}\right)\right)}
    K\left(m^{(-)}_{1,\ell_e}\left(\frac{\pi}{\ell_{E}}\right)\right)^{2}}{
    \ell_{e}^{2}}
    \label{eq:f_sigma_pi}
\end{equation}

Now let us assume that the two conditions \eqref{cond1_thm2}
stated in Theorem~\ref{thm2}
are satisfied and let us choose (for $M<E/2$ as is given in the condition of the theorem)
\begin{equation}
  \sigma_{e}=\begin{cases}
    1 & \text{for \ensuremath{e\le M},}\\
    -1 & \text{for \ensuremath{e\ge M+1}.}
  \end{cases}
\end{equation}
Then
\begin{equation}
  f_{\boldsymbol{\sigma}}(0)
  =2K\left(\frac{1}{2}\right)^{2}\left[\sum_{e=1}^{M}\frac{1}{\ell_{e}^{2}}-
    \sum_{e=M+1}^{E}\frac{1}{\ell_{e}^{2}}\right]
\end{equation}
 and the right inequality of
  \eqref{cond1_thm2} directly implies that
  $f_{\boldsymbol{\sigma}}(0)<0$.

  In order to prove the existence of the solution stated in
  Theorem~\ref{thm2}, it is left to show that
  \mbox{$f_{\boldsymbol{\sigma}}\left(\frac{\pi}{\ell_{E}}\right)>0$}, which would imply that
  $f_{\boldsymbol{\sigma}}$ vanishes for some $k\in(0, \frac{\pi}{\ell_{E}})$.
  Using our choice of signs and the identity
  $m^{(-)}_{1,\ell_e}(k)= m^{(-)}_{1,1}(k\ell_e)$
  we may rewrite Equation~(\ref{eq:f_sigma_pi}) as
\begin{equation}
  \begin{split}
    f_{\boldsymbol{\sigma}}\left(\frac{\pi}{\ell_{E}}\right)=
    & \sum_{e=1}^{M}
    \frac{4\sqrt{m^{(-)}_{1,1}\left(\frac{\pi\ell_{e}}{\ell_{E}}\right)
        \left(1-m^{(-)}_{1,1}\left(\frac{\pi\ell_{e}}{\ell_{E}}\right)\right)}
      K\left(m^{(-)}_{1,1}
        \left(\frac{\pi\ell_{e}}{\ell_{E}}\right)\right)^{2}}{
      \ell_{e}^{2}}\\
    - & \sum_{e=M+1}^{E-1}
    \frac{4\sqrt{m^{(-)}_{1,1}\left(\frac{\pi\ell_{e}}{\ell_{E}}\right)
        \left(1-m^{(-)}_{1,1}\left(\frac{\pi\ell_{e}}{\ell_{E}}\right)\right)}
      K\left(m^{(-)}_{1,1}\left(
          \frac{\pi\ell_{e}}{\ell_{E}}\right)\right)^{2}}{\ell_{e}^{2}}.
    \label{eq:fsigma}
  \end{split}
\end{equation}

  As $\sqrt{m(1-m)}K(m)^2$ is an increasing function for $m\in (0,\frac{1}{2})$
  and $m^{(-)}_{1,1}(k)$ is a decreasing function of its argument
  the left inequality in Equation~\eqref{cond1_thm2} implies
  \begin{equation}
    \sqrt{m_-
      \left(1-m_-\right)}
    K\left(m_-)\right)^{2}
    \sum_{e=M+1}^{E-1}\frac{1}{\ell_e^2}<
    \sqrt{m_+
      \left(1-m_+\right)}
    K\left(m_+)\right)^{2}
    \sum_{e=1}^M \frac{1}{\ell_e^2}
    \label{cond_implication}
  \end{equation}
  where
  $m_{+}=m^{(-)}_{1,1}\left(\frac{\pi\ell_{M}}{\ell_{1}}\right)$
  and $m_{-}=m^{(-)}_{1,1}\left(\frac{\pi\ell_{M+1}}{\ell_{1}}\right)$.
  The same monotonicity argument implies that
  the negative contributions in Equation~\eqref{eq:fsigma} are smaller than the
  left-hand side of inequality  Equation~\eqref{cond_implication}
  and that the positive contributions in
  Equation~\eqref{eq:fsigma} are larger than the
  right-hand side of inequality Equation~\eqref{cond_implication}.
  Thus Equation~\eqref{cond_implication} implies
  $f_{\boldsymbol{\sigma}}\left(\frac{\pi}{\ell_E}\right)>0$, as required.\hfill
\end{proof}

\section{Conclusions}
\label{sec:discussion}

We have established the existence of solutions of the stationary nonlinear
Schr{\"o}dinger equation on metric star graphs with a nodal point at the centre.
The existence is subject to certain conditions on the edge lengths
that can be satisfied for any numbers of edges $E\ge 3$.
We stress that some of these solutions are deep in the nonlinear regime where
finding any solutions is quite non-trivial.
Let us elaborate on that. The non-linear solutions come in one-parameter families, and a possible way to track those families (or curves) is to start from the solutions of the
corresponding linear Schr{\"o}dinger equation. Indeed, since the linear solutions
are good approximations for the nonlinear solutions with low intensities they may be used as
starting points for finding nonlinear solutions numerically. By slowly
changing parameters one may then find some spectral curves that extend
into the deep nonlinear regime.
However, there may be many solutions
on spectral curves that do not extend to arbitrary small intensities and
these are are much harder to find numerically. By focusing on solutions
which vanish at the centre our work shows how to construct such solutions.
These may then be used in a numerical approach to give a more complete
picture of spectral curves.

The solutions that we construct are characterised by their nodal count
structure. The nodal structure
on a spectral curve is constant
as long as the corresponding solutions do not vanish at the centre.
Our approach thus constructs the solutions where the nodal structure
changes along the corresponding spectral curve. In this way we have made
some progress in characterizing general solutions on star graphs in terms of their nodal
structure.

Many open questions remain. The main one being whether all spectral
curves of the NLS equation on a star graph may be found just by
combining the linear solutions with the non-linear solutions which
vanish at the centre. If not, how many other spectral curves remain
and
how can they be characterized?
Numerically we found that
apart from the ground state spectral curve it is generic for a~spectral curve
to have at least one point where the corresponding solution vanishes at
the centre. Of~course this leaves open how many spectral curves there are
where the corresponding solutions never vanishes at the centre.
Another interesting line of future research may be to extend some of our
results to tree graphs.

\subsection*{Acknowledgments}

R.B. and A.J.K. were supported by ISF (Grant No. 494/14).
S.G. would like to thank the Technion for hospitality
and the Joan and Reginald Coleman-Cohen Fund for support.

\appendix
\section{Elliptic Integrals and Jacobi Elliptic Functions}

\label{sec:elliptic}

We use the following definitions for elliptic integrals (the Jacobi
form) \begin{subequations}
\begin{align}
F(x|m):= & \int_{0}^{x}\frac{1}{\sqrt{1-u^{2}}\sqrt{1-m\,u^{2}}}du\\
K(m):= & F(1|m)\\
E(x|m):= & \int_{0}^{x}\frac{\sqrt{1-m\,u^{2}}}{\sqrt{1-u^{2}}}du\\
\Pi(x|a,m):= & \int_{0}^{x}\frac{1}{\sqrt{1-u^{2}}\sqrt{1-m\,u^{2}}(1-a\,u^{2})}du
\end{align}
\label{specfunc} 
\end{subequations} 
where $0\le x\le1$, $m\le1$
and $a\le1$. Note that our definition allows $m$ and $a$ to be
negative.

 The notation in the literature is far from being uniform. Our choice seems
the most concise for the present context and it is usually straight-forward to translate our definitions into the ones of any standard
reference on special functions. For instance, the \textit{NIST Handbook
of Mathematical Functions}~\cite{NIST} defines the three elliptical
integrals $F(\phi,k)$, $E(\phi,k)$ and $\Pi(\phi,\alpha,k)$ by
setting $x=\sin(\phi)$, $m=k^{2}$, and~$a=\alpha^{2}$ in our definitions
above.

 Jacobi's Elliptic function $\mathrm{sn}(x,m)$, the elliptic sine,
is defined as the inverse of $F(u|m)$
\begin{equation}
u=\mathrm{sn}(x,m)\qquad\Leftrightarrow\qquad x=F(u|m)\ .
\end{equation}

This defines $\mathrm{sn}(x,m)$ for $x\in[0,K(m)]$ which can straight-forwardly be extended to a periodic function with period $4K(m)$
by requiring $\mathrm{sn}(K(m)+x,m)=\mathrm{sn}(K(m)-x,m)$, $\mathrm{sn}(-x,m)=-\mathrm{sn}(x,m)$
and $\mathrm{sn}(x+4K(m),m)=\mathrm{sn}(x,m)$. The corresponding
elliptic cosine $\mathrm{cn}(x,m)$ is obtained by requiring that
it is a continuous function satisfying
\begin{equation}
\mathrm{cn}^{2}(x,m)+\mathrm{sn}^{2}(x,m)=1
\end{equation}
such that $\mathrm{cn}(0,m)=1$. It is useful to also define the non-negative
function
\vspace{6pt}
\begin{equation}
\mathrm{dn}(x,m):=\sqrt{1-m\,\mathrm{sn}^{2}(x,m)}.
\end{equation}

At $m=0$ and $m=1$ the elliptic functions can be expressed as \begin{subequations}
\begin{align}
\mathrm{sn}(x,0)= & \sin x, & \mathrm{sn}(x,1)= & \tanh x,\\
\mathrm{cn}(x,0)= & \cos x, & \mathrm{cn}(x,1)= & \cosh^{-1}x,\\
\mathrm{dn}(x,0)= & 1, & \mathrm{dn}(x,1)= & \cosh^{-1}x\ .
\end{align}
\end{subequations} 

Derivatives of elliptic functions can be expressed
in terms of elliptic functions \begin{subequations}
\begin{align}
\frac{d}{dx}\mathrm{sn}(x,m)= & \mathrm{cn}(x,m)\mathrm{dn}(x,m),\\
\frac{d}{dx}\mathrm{cn}(x,m)= & -\mathrm{sn}(x,m)\mathrm{dn}(x,m),\\
\frac{d}{dx}\mathrm{dn}(x,m)= & -m\,\mathrm{sn}(x,m)\mathrm{cn}(x,m)\ .
\end{align}
\end{subequations} 

The first of these equations implies that $u=\mathrm{sn}(x,m)$
is a solution of the first order ordinary differential equation
\begin{equation}
\frac{du}{dx}=\sqrt{1-u^{2}}\sqrt{1-mu^{2}}\ .\label{eq:sn_ODE}
\end{equation}

\end{document}